\documentclass[fleqn,usenatbib]{mnras}

\usepackage{newtxtext,newtxmath}

\usepackage[T1]{fontenc}

\DeclareRobustCommand{\VAN}[3]{#2}
\let\VANthebibliography\thebibliography
\def\thebibliography{\DeclareRobustCommand{\VAN}[3]{##3}\VANthebibliography}

\usepackage{graphicx}	
\usepackage{amsmath}	

\title[]{Chromospheric turbulence as a regulator of stellar wind mass flux}

\author[M. Shoda et al.]{
Munehito Shoda,$^{1}$\thanks{E-mail: shoda.m.astroph@gmail.com}
Tom Van Doorsselaere$^{2}$
and Allan Sacha Brun$^{3,4}$
\\
$^{1}$Department of Earth and Planetary Science, School of Science, The University of Tokyo, 7-3-1 Hongo, Bunkyo-ku, Tokyo 113-0033, Japan\\
$^{2}$Centre for mathematical Plasma Astrophysics, Department of Mathematics, KU Leuven, Celestijnenlaan 200B, 3001 Leuven, Belgium\\
$^{3}$Universit\'e Paris-Saclay, Universit\'e Paris Cit\'e, CEA, CNRS, AIM, 91191, Gif-sur-Yvette, France\\
$^{4}$Institute for Space-Earth Environmental Research, Nagoya University, Furo-cho, Chikusa-ku, Nagoya, Aichi 464-8601, Japan
}

\date{Accepted XXX. Received YYY; in original form ZZZ}

\pubyear{\the\year{}}

\begin{document}
\label{firstpage}
\pagerange{\pageref{firstpage}--\pageref{lastpage}}
\maketitle

\begin{abstract}
The mass flux of solar and stellar winds is a key quantity for stellar evolution and space weather, yet its physical regulation mechanism remains an unsolved problem. In particular, conventional Alfv\'en wave--driven models that self-consistently connect the stellar surface to the stellar wind fail to reproduce the observed scaling between stellar X-ray flux and mass-loss rate, a discrepancy that can be largely attributed to the dissipation of a substantial fraction of the wave energy by chromospheric turbulence. To address this issue, we aim to clarify the role of chromospheric turbulence in regulating the stellar wind mass flux. We perform one-dimensional wave-driven wind simulations, comparing cases with and without chromospheric turbulence suppression to assess its impact on coronal and wind properties. We find that suppressing chromospheric turbulence leads to a systematic increase in the coronal particle flux, and hence the wind mass flux, by up to an order of magnitude, particularly in regions of moderately strong magnetic field. This behavior arises from a combination of changes in the Poynting flux at the coronal base and in the asymptotic wind speed. Furthermore, the model with chromospheric turbulence suppression reproduces the observed empirical scaling between coronal magnetic field strength and mass flux without invoking additional energy input mechanisms such as interchange reconnection. These results identify the chromospheric turbulence as a key factor in regulating stellar wind mass flux and highlight the importance of incorporating its effects in models that connect the stellar surface and the stellar wind.
\end{abstract}

\begin{keywords}
solar wind -- Sun: chromosphere -- waves -- turbulence
\end{keywords}



\section{Introduction}

Stellar winds are steady outflows from stars \citep{Lamers_1999_book}, which affect the evolution of the universe in various ways. Stellar winds regulate the exchange of mass and metals between stars and the interstellar medium, thereby influencing galaxy evolution \citep{Hofner_2018_AARev}. In stellar evolution, stellar wind plays a key role by driving mass and angular momentum loss \citep{Brott_2011_AandA, Gallet_2013_AA, Gallet_2015_AA}. For stars hosting a planetary systems, stellar winds can cause atmospheric escape \citep{Dong_2017_ApJ, Rodriguez_Monoz_2019_AandA,  Canet_2024_MNRAS}, profoundly impacting planetary atmospheres and surface environments. 

One of the most critical issues in stellar wind research is understanding how the mass flux (or mass-loss rate) evolves over time. While the spectroscopic signatures of stellar winds are observable in massive stars \citep{Kudritzki_2000_ARAA} and possibly in low-mass pre-main-sequence stars \citep{Edwards_2006_ApJ, Kuwan_2007_ApJ, Erkal_2022_AandA}, detecting winds from low-mass main-sequence stars remains challenging due to the absence of direct observational signatures. Nevertheless, various indirect methods have been proposed \citep{Wood_2001_ApJ, Wood_2002_ApJ, Jardine_2019_MNRAS, Kislyakova_2024_NatAs}, progressively increasing the number of available samples and revealing a positive correlation between the mass-loss rate ($\dot{M}_w$) and X-ray luminosity ($L_{\rm X}$) \citep{Wood_2021_ApJ, Vidotto_2021_LRSP}. However, some stars \citep{Wood_2005_ApJ, Wood_2014_ApJ}, as well as the Sun in different activity phases \citep{Cohen_2011_MNRAS, Shoda_2023_ApJ}, deviate from this trend, leaving the reliability of the $L_{\rm X}$-$\dot{M}_w$ relation uncertain. The existence of such outliers suggests that the empirical $L_{\rm X}$–$\dot{M}_w$ relation may not be inherently robust. This motivates a theoretical investigation into whether this relation can arise from first principles, and what physical mechanisms are responsible for producing it. From this standpoint, theoretical modelling constitutes an indispensable approach to understanding stellar winds in low-mass stars.

Numerous studies have attempted to reveal the physical properties of stellar winds using theoretical models originally developed for the solar wind. A prominent example is the application of the AWSoM model \citep{Sokolov_2013_ApJ, van_der_Holst_2014_ApJ}, a global solar wind model based on Alfv\'en-wave heating and acceleration, to stellar winds \citep{Alvarado-Gomez_2016_AA, Pognan_2018_ApJ, Airapetian_2021_ApJ, Evensberget_2021_MNRAS, Evensberget_2022_MNRAS}. However, the AWSoM model assumes an artificial scaling in which the injected energy flux at the boundary is proportional to the magnetic field strength. While this assumption is found to work in reproducing the solar atmosphere \citep{van_der_Holst_2014_ApJ, Sachdeva_2019_ApJ}, its validity for much younger stars or stars of different spectral types remains uncertain \citep{Garraffo_2016_ApJ, Airapetian_2017_ApJ, Vidotto_2023_A&A}. Since the mass-loss rate is nearly proportional to the boundary energy flux \citep{Boro_Saikia_2020_AandA, Kavanagh_2021_MNRAS}, this uncertainty directly leads to a corresponding uncertainty in the mass-loss rate.

To reduce the uncertainties in boundary parameters, it is effective to adopt models that couple the stellar surface (photosphere) with the stellar wind. This is because physical quantities at the photosphere can be directly inferred from spectroscopy \citep{Passenger_2018_AandA, Kochukhov_2020_AA, Jahandar_2025_ApJ} and are generally less uncertain than those in the upper atmosphere. Numerous one-dimensional models have been developed for connecting the photosphere and solar wind \citep{Suzuki_2005_ApJ, Cranmer_2007_ApJ, Shoda_2018_ApJ_a_self-consistent_model, Shoda_2022_ApJ, Shimizu_2022_ApJ}, and more recently, three-dimensional models have emerged \citep{Matsumoto_2021_MNRAS, Iijima_2023_ApJ}. Nonetheless, one-dimensional models remain suitable for wide-range parameter surveys required in stellar wind studies. Several such models have estimated the mass-loss rate and its dependence on stellar parameters \citep{Cranmer_2011_ApJ, Suzuki_2013_PASJ, Suzuki_2018_PASJ, Shoda_2020_ApJ, Sakaue_2021_ApJ}. \citet{Shoda_2020_ApJ} developed an advanced wave-driven wind model incorporating plasma compressibility and Alfv\'en-wave turbulence. That model, however, does not reproduce the observed increase in mass-loss rate with stellar rotation or X-ray luminosity, because the enhanced wave dissipation in the chromosphere at high rotation limits the energy reaching the corona.

A possible solution to the insufficient energy injection by Alfvén waves is to incorporate the effect of interchange reconnection—a magnetic reconnection between open and closed field lines—which has attracted significant attention in the solar wind studies \citep{Fisk_2003_JGR, Antiochos_2011_ApJ, Bale_2023_Nature, Iijima_2023_ApJ, Wang_2024_SoPh}. \citet{Shoda_2023_ApJ} suggested that introducing this effect phenomenologically could potentially explain the observed relationship between mass-loss rate and X-ray luminosity. However, their approach is based on solar-specific phenomenology \citep{Wang_2020_ApJ} and may not be directly applicable to other stars, rendering their results uncertain. 

In this work, we take a different approach and examine whether modifying the treatment of chromospheric turbulent dissipation allows an Alfv\'en wave--driven framework to reproduce the observed $L_{\rm X}$--$\dot{M}_w$ relation. Previous Alfv\'en-wave turbulence models that connect the photosphere to the corona and the solar wind have typically assumed that waves are generated by the incoherent flows on spatial scales smaller than individual photospheric magnetic elements, as illustrated in Figure~1 of \citet{van_Ballegooijen_2011_ApJ}, and that turbulence develops independently within individual flux tubes \citep{Cranmer_2007_ApJ, Chandran_2019_JPP, Shoda_2020_ApJ}. Recent radiation-MHD simulations, however, support a different picture: waves are predominantly generated by the rotational motion of flux tubes \citep{Finley_2022_AA, Breu_2022_AA, Kuniyoshi_2023_ApJ}, and magnetic dissipation occurs preferentially at the boundaries between neighboring flux tubes rather than within their interiors \citep{van_Ballegooijen_2017_ApJ_coronal_loop,Kannan_2024_MNRAS, Breu_2026_MNRAS}. This scenario suggests that turbulent dissipation is weak in the photosphere and chromosphere where adjacent flux tubes remain unmerged or have only recently merged. These findings motivate a reassessment of previous wave-driven wind models through a revised treatment of chromospheric turbulence.


In light of the discussion above, we investigate how the solar-wind mass flux depends on the efficiency of chromospheric turbulent dissipation, and whether suppressing this dissipation leads to improved agreement with observational constraints. To this end, we perform a series of one-dimensional solar-wind simulations with and without chromospheric turbulence dissipation. To assess the observational relevance, we compare our results with observational constraints on the relation between the coronal magnetic-field strength and the upward particle flux \citep{Wang_2020_ApJ, Stansby_2021_AandA}. This scaling provides a quantitative benchmark for evaluating whether the turbulence-suppressed model reproduces a more realistic magnetic-field dependence of the solar-wind mass flux within the framework of Alfv\'en-wave-driven wind models.

\section{Model description}

\subsection{Basic equations}

We model the solar wind along a static, open flux tube using a one-dimensional MHD framework that includes gravity, thermal conduction, and radiative cooling. For simplicity, a thin vertical flux tube is assumed, with the background magnetic field nearly radial. The governing equations are given below.
\begin{align}
    \frac{\partial}{\partial t} \rho + \frac{1}{r^2 f} \frac{\partial}{\partial r} \left( \rho v_r r^2 f \right) = 0, \label{eq:basic_mass_cons}
\end{align}
\begin{align}
    \frac{\partial}{\partial t} \left( \rho v_r \right) &+ \frac{1}{r^2 f} \frac{\partial}{\partial r} \left[ \left( \rho v_r^2 + p_{\rm T}  \right) r^2 f \right] \notag \\
    &= \left( p + \frac{1}{2} \rho \boldsymbol{v}_\perp^2 \right) \frac{d}{dr} \ln \left( r^2 f \right) - \rho \frac{GM_\odot}{r^2},
\end{align}
\begin{align}
    \frac{\partial}{\partial t} \left( \rho \boldsymbol{v}_\perp \right) &+ \frac{1}{r^2 f} \frac{\partial}{\partial r} \left[ \left( \rho v_r \boldsymbol{v}_\perp - \frac{B_r \boldsymbol{B}_\perp}{4 \pi} \right) r^2 f \right] \notag \\
    &= - \frac{1}{2} \left( \rho v_r \boldsymbol{v}_\perp - \frac{B_r \boldsymbol{B}_\perp}{4 \pi} \right) \frac{d}{dr} \ln \left( r^2 f \right) + \rho \boldsymbol{D}_\perp^v, \label{eq:basic_transverse_eom}
\end{align}
\begin{align}
    \frac{d}{dr} \left( B_r r^2 f \right) = 0, \label{eq:magnetic_flux_conservation}
\end{align}
\begin{align}
    \frac{\partial}{\partial t} \boldsymbol{B}_\perp &+ \frac{1}{r^2 f} \frac{\partial}{\partial r} \left[ \left( v_r \boldsymbol{B}_\perp - B_r \boldsymbol{v}_\perp \right) r^2 f \right] \notag \\
    &= \frac{1}{2} \left( v_r \boldsymbol{B}_\perp - B_r \boldsymbol{v}_\perp \right) \frac{d}{dr} \ln \left( r^2 f \right) + \sqrt{4 \pi \rho} \boldsymbol{D}_\perp^b, \label{eq:basic_induction_equation}
\end{align}
\begin{align}
    \frac{\partial}{\partial t} e &+ \frac{1}{r^2 f} \frac{\partial}{\partial r} \left[ \left( \left( e+ p_{\rm T} \right) v_r - \frac{B_r}{4 \pi} \left( \boldsymbol{v}_\perp \cdot \boldsymbol{B}_\perp \right) \right) r^2 f \right] \notag \\
    &= Q_{\rm cnd} + Q_{\rm rad} - \rho v_r \frac{GM_\odot}{r^2},
\end{align}
where $\boldsymbol{v}_\perp$ and $\boldsymbol{B}_\perp$ represent the perpendicular components of the veloctiy and magnetic field, respectively. $e$ and $p_{\rm T}$ denote the total energy density and total pressure, respectively, defined by
\begin{align}
    e = e_{\rm int}+ \frac{1}{2} \rho \boldsymbol{v}^2 + \frac{\boldsymbol{B}_\perp^2}{8\pi}, \ \ \ p_{\rm T} = p + \frac{\boldsymbol{B}_\perp^2}{8 \pi},
\end{align}
where $ e_{\rm int}$ denotes the internal energy per unit volume \citep{Shoda_2021_AA}

$f$ represents the degree to which a magnetic flux tube expands beyond a purely radial expansion, known as the super-radial expansion factor. Assuming a static flux tube, we treat $f$ as a time-independent function of radial distance. The profile $f(r)$ is constructed by combining the coronal flux tube model derived from PFSS extrapolation with a chromospheric flux tube model (see Sections~\ref{sec:pfss_field_tracing} and~\ref{sec:combining_chromospheric_flux_tube}). 

The terms $Q_{\rm cnd}$ and $Q_{\rm rad}$ in the energy equation represent heating by thermal conduction and radiation, respectively, and are essential for the formation of the transition region and the solar wind. Thermal conduction is modelled using a Spitzer--Harm-type heat flux \citep{Spitzer_1953_PhysRev}, with a correction that gradually reduces the conductive flux for $r > r_{\rm cnd}$. The conductive heating term is written as
\begin{align}
    Q_{\rm cnd} &= - \frac{1}{r^2 f} \frac{\partial}{\partial r} \left( r^2 f \, q_{\rm cnd} \right), \\
    q_{\rm cnd} &= - \left( \frac{r_{\rm cnd}}{r} \right)^2 \kappa_{\rm SH} \, T^{5/2} \frac{\partial T}{\partial r},
\end{align}
where $r_{\rm cnd}/R_\odot = 10$, and $\kappa_{\rm SH}$ is set to $10^{-6}$ in cgs units. Radiative cooling includes an exponential term near the photosphere to mimic optically thick radiation, and in the upper atmosphere, it follows an optically thin formulation using a radiative loss function based on the CHIANTI atomic database version 10 \citep{Dere_1997_AA, Del_Zanna_2021_ApJ}, extended to the chromospheric temperature \citep{Goodman_2012_ApJ, Iijima_2016_PhD}. The detailed formulation of the radiative cooling term is described in Section 2.5 of \citet{Shoda_2023_ApJ}.

$\boldsymbol{D}_\perp^v$ and $\boldsymbol{D}_\perp^b$ are phenomenological terms representing the turbulent dissipation of Alfv\'en waves, as described in the following section. From Equations \eqref{eq:basic_mass_cons}, \eqref{eq:basic_transverse_eom}, and \eqref{eq:basic_induction_equation}, it is straightforward to derive the following Alfv\'en-wave energy conservation law:
\begin{equation}
    \frac{\partial}{\partial t} e_{\rm A} + \frac{1}{r^2 f} \frac{\partial}{\partial r} \left( F_{\rm A} r^2f \right) = - Q_{\rm work} - Q_{\rm turb},
\end{equation}
where $e_{\rm A}$ and $F_{\rm A}$ denote the energy density and energy flux of the Alfv'en wave, respectively, and are given by
\begin{equation}
    e_{\rm A} = \frac{1}{2}\rho v_\perp^2 + \frac{B_\perp^2}{8 \pi}, \ \ \ \ F_{\rm A} = \left( e_{\rm A} + \frac{B_\perp^2}{8\pi} \right) v_r - \frac{B_r}{4 \pi} \left( \boldsymbol{v}_\perp \cdot \boldsymbol{B}_\perp \right). \label{eq:alfven_wave_energy_density_flux}
\end{equation}
$Q_{\rm work}$ and $Q_{\rm turb}$ represent the wave energy losses associated with wave–pressure work and turbulent dissipation, respectively, and are expressed as
\begin{align}
    Q_{\rm work} &= - v_r \frac{\partial}{\partial r} \left( \frac{B_\perp^2}{8 \pi} \right) + \left(\rho v_\perp^2 - \frac{B_\perp^2}{4 \pi} \right) \frac{v_r}{2} \frac{d}{dr} \ln \left(r^2 f\right), \\
    Q_{\rm turb} &= - \rho \left( \boldsymbol{v}_\perp \cdot \boldsymbol{D}_\perp^v + \frac{\boldsymbol{B}_\perp}{\sqrt{4 \pi \rho}} \cdot \boldsymbol{D}_\perp^b \right). \label{eq:turbulent_dissipation_general_form}
\end{align}
The term $Q_{\rm work}$ includes contributions from both the nonlinear generation of magnetoacoustic waves driven by local fluctuations in the wave pressure \citep{Hollweg_1982_SolPhys, Kudoh_1999_ApJ, Shoda_2020_ApJ, Shimizu_2022_ApJ} and the large-scale acceleration of the solar wind resulting from the global wave–pressure gradient \citep{Alazraki_1971_AA, Belcher_1971_ApJ}.

\subsection{Turbulence modelling}

The terms $\boldsymbol{D}_\perp^v$ and $\boldsymbol{D}_\perp^b$ in Equations \eqref{eq:basic_transverse_eom} and \eqref{eq:basic_induction_equation} are phenomenological terms designed to account for turbulent dissipation within a one-dimensional framework and are given as follows \citep{Shoda_2018_ApJ_a_self-consistent_model, Shoda_2021_AA}.
\begin{align}
    D_{x,y}^v &= - \frac{c_{\rm dis}}{4\lambda_\perp} \left( z_\perp^+ z_{x,y}^- +  z_\perp^- z_{x,y}^+ \right), \label{eq:phenomenological_awt_vsource} \\
    D_{x,y}^b &= - \frac{c_{\rm dis}}{4\lambda_\perp} \left( z_\perp^+ z_{x,y}^- -  z_\perp^- z_{x,y}^+ \right), \label{eq:phenomenological_awt_bsource}
\end{align}
where $c_{\rm dis}$ is a dimensionless measure of dissipation efficiency, and $\lambda_\perp$ is the correlation length perpendicular to the mean magnetic field. The Els\"asser variables $\boldsymbol{z}^\pm_\perp$ are defined as follows.
\begin{equation}
    z_{x,y}^\pm = v_{x,y} \mp \frac{B_{x,y}}{\sqrt{4 \pi \rho}}, \ \ \ \ z_\perp^\pm = \sqrt{{z_x^\pm}^2+{z_y^\pm}^2}. \label{eq:elsasser_definition}
\end{equation}
We note that the phenomenological terms are slightly modified from the original form in \citet{Shoda_2018_ApJ_a_self-consistent_model}. It follows directly from Equations \eqref{eq:turbulent_dissipation_general_form}, \eqref{eq:phenomenological_awt_vsource}, and \eqref{eq:phenomenological_awt_bsource} that the turbulent dissipation rate is given by
\begin{align}
    Q_{\rm turb} = c_{\rm dis} \rho \frac{z_\perp^+ \left( z_\perp^- \right)^2 + z_\perp^- \left( z_\perp^+ \right)^2}{4 \lambda_\perp}, \label{eq:Qturb_elsasser_form}
\end{align}
in agreement with the mean-field description of reduced-MHD Alfv\'en-wave turbulence \citep{Dmitruk_2002_ApJ, Verdini_2007_ApJ, Lionello_2014_ApJ, Downs_2016_ApJ}.

The perpendicular correlation length $\lambda_\perp$ is assumed to scale with the flux-tube radius \citep{Hollweg_1986_JGR}. From magnetic flux conservation, it follows that
\begin{equation}
    \lambda_\perp = \lambda_{\perp,\odot} \sqrt{\frac{B_{r,\odot}}{B_r}} ,
\end{equation}
where the photospheric correlation length is set to $\lambda_{\perp,\odot} = 150$ km \citep{Utz_2009_AA, Shoda_2024_AA}, and the photospheric radial field $B_{r,\odot}$ is specified in Section~\ref{sec:combining_chromospheric_flux_tube}.

The suppression of chromospheric turbulence is implemented by adjusting the coefficient $c_{\rm dis}$. When turbulence is not suppressed, $c_{\rm dis}$ is treated as a constant independent of space and time,
\begin{align}
    c_{\rm dis} = 0.1,
\end{align}
a value supported by theoretical and observational studies in the literature \citep{van_Ballegooijen_2016_ApJ, Verdini_2019_SolPhys}. When suppression is applied, noting that temperature is the simplest criterion distinguishing the chromosphere from the corona, we introduce a temperature-dependent quenching of $c_{\rm dis}$:
\begin{align}
    c_{\rm dis} = 0.1 \times \min \left[ 1, \ \max \left[ 0, \ \log \left( \frac{T}{10^4 {\rm \ K}}\right) \right] \right],
\end{align}
so that turbulent dissipation vanishes for $T<10^4 {\rm \ K}$ and recovers to the standard value for $T>10^5 {\rm \ K}$.

\subsection{Field-line tracing with magnetic field extrapolation \label{sec:pfss_field_tracing}}
Open flux tubes exhibit diverse geometries that contribute to the variability of the solar wind \citep{Wang_1990_ApJ, Arge_2000_JGR, Dakeyo_2024_AandA, Tokoro_2026_ApJ}. Although these structures are often modeled using analytical functions with limited free parameters \citep{Kopp_1976_SolPhys}, real open flux tubes are not necessarily well represented by such forms. Therefore, instead of adopting analytical expressions, this study estimates the flux tube geometry from magnetic field extrapolation and uses it as input for a one-dimensional model.

We use the Potential Field Source Surface (PFSS) method \citep{Altschuler_1969_SolPhys, Schatten_1969_SolPhys} to extrapolate the coronal magnetic field, which depends on the input magnetogram and source surface height. During solar minimum, the coronal field structure is sensitive to the polar field, which is difficult to observe. We therefore adopt the ADAPT model \citep{Worden_2000_SoPh, Arge_2010_AIPC, Arge_2013_AIPC, Hickmann_2015_SoPh}, which estimates the polar field from physical modeling, as the input magnetogram. Since the mass flux scaling law used for comparison is based on 1998–2011 data, we use ADAPT-KPVT and ADAPT-VSM data from the same period \citep[collectively referred to as ADAPT-NSO following][]{Wallace_2019_SoPh}. The source surface height is adjusted so that the heliospheric open flux from the PFSS model matches that derived from the in-situ data \citep{Shoda_2025_ApJ}.

\begin{table}
	\centering
	\caption{Summary of the sampling of magnetic flux tubes used in this work. The first column lists the Carrington Rotation number and the corresponding central date. The second column gives the source surface height in units of the solar radius adopted for the PFSS extrapolation. The third column indicates the number of flux tubes extracted from each Carrington Rotation.\label{table:cr_selection}}
	\label{tab:example_table}
	\begin{tabular}{rrrr} 
		\hline
		CR number (mid. date) & \begin{tabular}{r} source-surface \hspace{-1em} \\ height [$R_\odot$] \hspace{-1em} \end{tabular} & \begin{tabular}{r} number of traced \hspace{-1em} \\ field lines \hspace{-1em} \end{tabular} \\
		\hline
		1933 (04/03/1998) & 1.942 & 4 \\
		1947 (21/03/1999) & 1.954 & 5 \\
		1961 (06/04/2000) & 2.190 & 6 \\
        1968 (13/10/2000) & 2.034 & 3 \\
        1975 (23/04/2001) & 1.927 & 4 \\
        1982 (30/10/2001) & 1.839 & 4 \\
        2003 (26/05/2003) & 1.884 & 4 \\
        2031 (28/06/2005) & 1.963 & 5 \\
        2045 (15/07/2006) & 2.470 & 4 \\
        2058 (04/07/2007) & 1.623 & 4 \\
        2073 (16/08/2008) & 2.472 & 4 \\
        2087 (02/09/2009) & 1.616 & 4 \\
        2101 (19/09/2010) & 1.686 & 3 \\
        2115 (06/10/2011) & 1.455 & 4 \\
		\hline
	\end{tabular}
\end{table}

The procedure of field-line tracing is as follows. We first select a Carrington Rotation (CR) between 1998 and 2011. Using the magnetic field map corresponding to the central date of the selected CR, obtained from ADAPT-NSO, we perform the PFSS extrapolation with an appropriate source-surface height to obtain the three-dimensional magnetic field data. Random points are then selected on the source surface, and field lines are traced back to the photosphere to determine their geometry. This procedure is repeated for a sufficient number of CRs in 1998-2011. Table~\ref{table:cr_selection} summarizes the selected CRs, the corresponding source-surface heights, and the number of traced field lines in each CR. We used the open-source software {\it pfsspy} \citep{Stansby_2020_JOSS} for both PFSS extrapolation and field-line tracing.

\subsection{Incorporating flux-tube expansion in the chromosphere \label{sec:combining_chromospheric_flux_tube}}

Although the PFSS method provides the magnetic field structure from the photosphere to the heliosphere, its results are only valid above the transition region where the plasma beta is sufficiently low, indicating that the magnetic configuration within the chromosphere is not reliably captured. In fact, magnetic fields traced back to the photosphere using PFSS often have strengths of 1–10 G \citep{Fujiki_2015_SoPh}, whereas photospheric fields are typically concentrated in magnetic bright points with strengths of 100–1000 G \citep{Berger_2001_ApJ, Tsuneta_2008_ApJ, Utz_2013_AandA}, suggesting that the PFSS values are too low. This discrepancy arises because PFSS model does not capture chromospheric magnetic structures, in particular the flux-tube expansion that reduces the field strength \citep{Bruls_1995_AandA, Gu_1997_ApJ, Cranmer_2005_ApJ, Ishikawa_2021_ScienceAdvances}. For this reason, we independently model the chromospheric magnetic field and smoothly connect it with the PFSS solution to construct a consistent large-scale field-aligned structure.

We assume that photospheric magnetic concentrations have near-equipartition strength, with magnetic pressure balancing gas pressure, and that flux tubes expand vertically to maintain this relation until they merge with neighboring tubes \citep{Cranmer_2005_ApJ}. Let $H_\odot$ be the pressure scale height at the photosphere; then, the magnetic field strength in the chromosphere can be approximated as \citep[see also][]{van_Ballegooijen_2011_ApJ, Chandran_2019_JPP}:
\begin{align}
    B_r (r) \approx B_{r,\odot} \exp \left( - \frac{r-R_\odot}{2 H_\odot} \right), \label{eq:magnetic_field_chromosphere_model}
\end{align}
where $B_{r,\odot}$ denotes the field strength at the photosphere. In this study, we adopt $H_\odot = 1.74 \times 10^7$ km and $B_{r,\odot}=1300~\mathrm{G}$, where the latter represents a typical value close to the equipartition field strength at the photosphere.

The chromospheric magnetic field model is combined with the PFSS magnetic field model as described below. We denote the radial field strength derived from the PFSS method as $B_r^{\rm PFSS} (r)$. Since the field-line tracing is limited to the region between the photosphere ($r = R_\odot$) and the source surface ($r = R_{\rm SS}$), we extrapolate $B_r^{\rm PFSS}(r)$ beyond the source surface ($r \ge R_{\rm SS}$) as $B_r^{\rm PFSS}(r) \propto r^{-2}$, following the definition of source surface. First, the radial magnetic field strength in the coronal base, which serves as the interface between the PFSS field and the chromospheric field, is calculated using the PFSS extrapolation data as follows.
\begin{align}
    B_{r,{\rm cb}} = B_r^{\rm PFSS} \left( r_{\rm cb} \right),
\end{align}
where $r_{\rm cb}$ denotes the radial distance of the coronal base, with $r_{\rm cb}/R_\odot$ ranging from 1.003 to 1.01 depending on the result of the field-line tracing. The flux tube expansion in the chromosphere and corona are then modeled accordingly:
\begin{align}
    f_{\rm chr} (r) = \frac{\eta_{\rm exp}}{\sqrt{\eta_{\rm exp}^2-1}} \exp\left( \frac{r-R_\odot}{2 H_\odot}\right),
\end{align}
\begin{align}
    f_{\rm cor} (r) = \frac{r_{\rm cb}^2 B_{r,{\rm cb}}}{r^2 B^{\rm PFSS} (r)} \ \left( r \ge r_{\rm cb} \right), \ \ \ f_{\rm cor} (r) = 1 \ \left( r<r_{\rm cb} \right),
\end{align}
where $\eta_{\rm exp} = R_\odot^2 B_{r,\odot}/( r_{\rm cb}^2 B_{r, {\rm cb}} )$ denotes the total flux-tube expansion in the chromosphere. We note that $f_{\rm exp} = \lim_{r \to \infty} f_{\rm cor} (r)$ is the conventional (coronal) expansion factor, which is known to be anti-correlated with the solar wind speed \citep{Wang_1990_ApJ, Arge_2000_JGR}. 

\begin{figure}
    \centering
    \includegraphics[width=0.85\columnwidth]{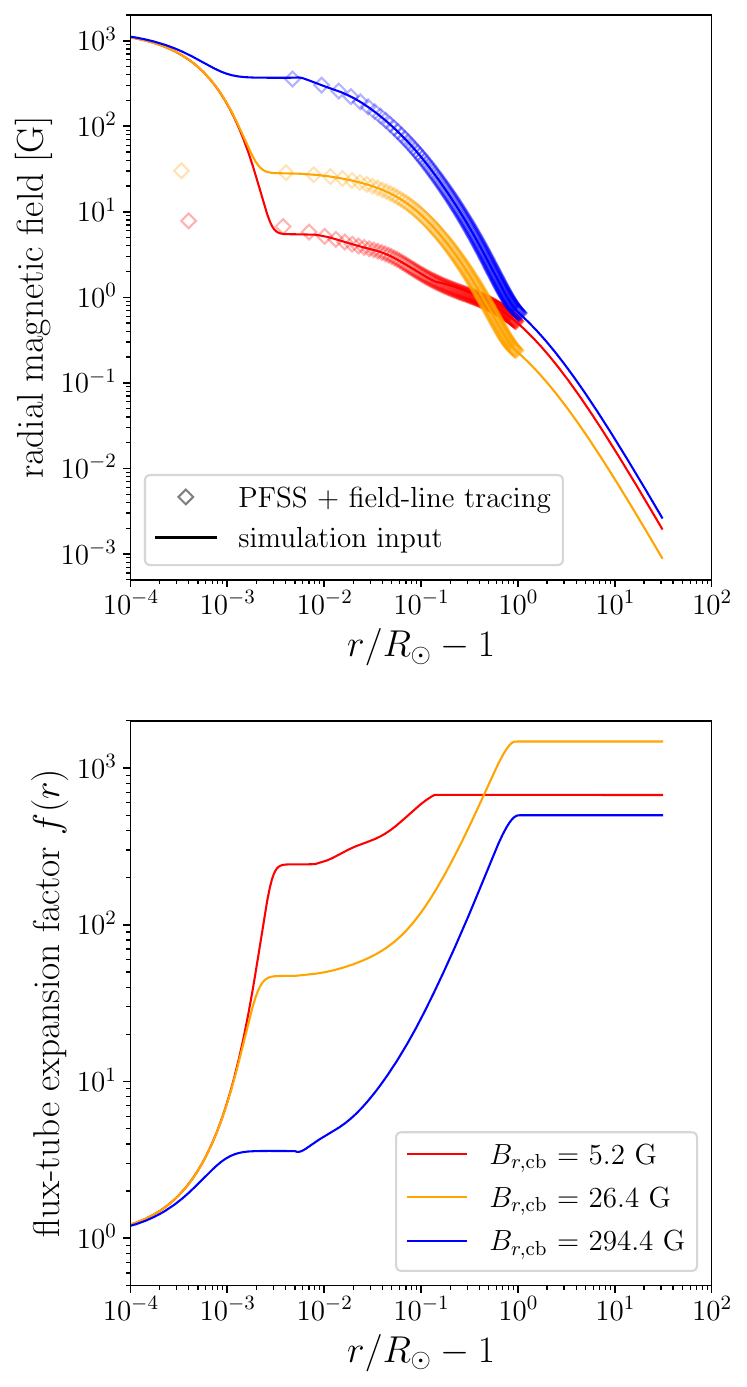}
    \caption{Top panel: Comparison between the radial magnetic field $B_r$ distribution derived from the PFSS extrapolation (diamonds) and that used as input in the simulation (solid lines). Different colors represent different magnetic field lines. Bottom panel: Radial distribution of the corresponding flux-tube expansion factor. The color coding is the same as in the top panel.
    \label{fig:extrapolated_magnetic_field_examples}}
\end{figure}

The profile of $f(r)$ is given by the combination of the two functions as follows.
\begin{align}
    f (r) = \frac{\eta_{\rm exp }f_{\rm chr}(r) f_{\rm cor} (r)}{\sqrt{f_{\rm chr} (r)^2 + \eta_{\rm exp}^2 f_{\rm cor} (r)^2}}. \label{eq:expansion_factor_combined}
\end{align}
We note that $f_\odot = f(R_\odot) = 1$. 

\begin{figure}
    \centering
    \includegraphics[width=0.85\columnwidth]{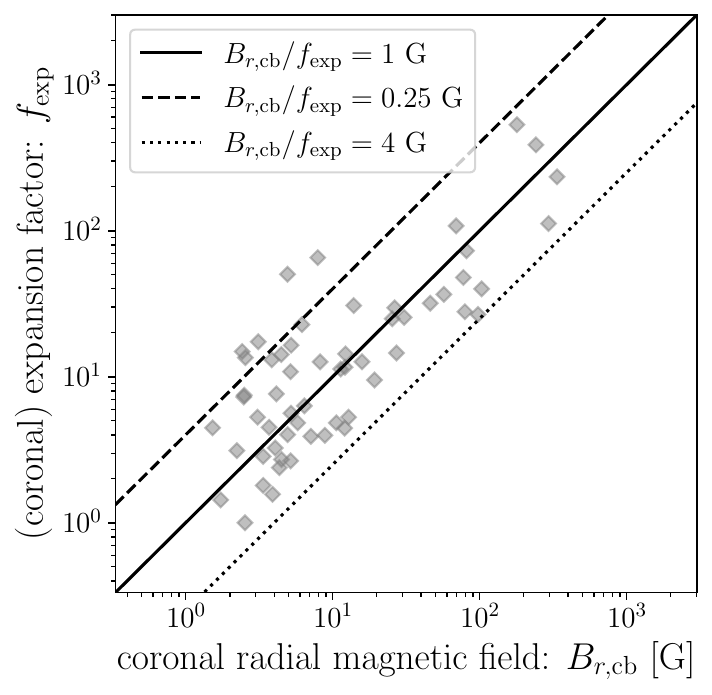}
    \caption{Scatter plot of the coronal magnetic field strength ($B_{r,{\rm cb}}$, horizontal axis) versus the coronal flux-tube expansion factor ($f_{\rm exp} = \lim_{r \to \infty} f_{\rm cor} (r)$, vertical axis) used as simulation input in this study. Each symbol represents an individual flux tube. The three lines indicate contours of constant $B_{r,{\rm cb}}/f_{\rm exp}$, with the solid, dashed, and dotted lines corresponding to $B_{r,{\rm cb}}/f_{\rm exp} = 1~{\rm G}$, $0.25~{\rm G}$, and $4~{\rm G}$, respectively.
    \label{fig:extrapolated_magnetic_field_statistics}}
\end{figure}

Combining Equation~\eqref{eq:expansion_factor_combined} with the magnetic flux conservation law (Equation~\eqref{eq:magnetic_flux_conservation}) yields the profile of $B_r(r)$, which can be straightforwardly shown to be consistent with chromospheric field models and PFSS fields. Given that $\eta_{\rm exp}^2$ is typically much greater than unity, $f(r)$ in the chromosphere ($r \approx R_\odot$) is approximated as
\begin{align}
    f(r) \approx e^{(r-R_\odot) / (2H_\odot)} \ \ \left( r \approx R_\odot \right),
\end{align}
leading to
\begin{align}
    B_r(r) = \frac{B_{r,\odot} R_\odot^2 f_\odot}{r^2 f(r)} \approx B_{r,\odot} \exp \left( - \frac{r-R_\odot}{2 H_\odot} \right) \ \ \left( r \approx R_\odot \right),
\end{align}
which is consistent with Equation~\eqref{eq:magnetic_field_chromosphere_model}. Meanwhile, above the corona ($r \ge r_{\rm cb}$), where $f_{\rm chr}(r) \gg \eta_{\rm exp} f_{\rm cor}(r)$ typically holds, we obtain
\begin{align}
    f(r) \approx \eta_{\rm exp} f_{\rm cor}(r) \ \ \left( r \ge r_{\rm cb} \right).
\end{align}
Applying the magnetic flux conservation then gives
\begin{align}
    B_r(r) = \frac{B_{r,\odot} R_\odot^2 f_\odot}{r^2 f(r)} \approx \frac{B_{r,\odot} R_\odot^2}{r^2 \eta_{\rm exp} f_{\rm cor}(r)} = B_r^{\rm PFSS} (r) \ \ \left( r \ge r_{\rm cb} \right),
\end{align}
which shows the consistency with the radial field from the PFSS extrapolation.

Figure~\ref{fig:extrapolated_magnetic_field_examples} presents examples of the radial magnetic field profile ($B_r$, top panel) and the flux-tube expansion factor ($f(r)$, bottom panel) obtained from the PFSS extrapolation (diamonds), together with the corresponding input profiles constructed using a chromospheric model. As shown in the top panel, the input and PFSS-derived $B_r$ profiles show good agreement in the coronal region, while the simulation input additionally captures the weakening of the magnetic field in the chromosphere, which is not represented in the PFSS solution. We performed the same comparison for all cases and confirmed that the adopted $B_r(r)$ profiles provide a reliable representation of the PFSS-extrapolated magnetic field.

As summarized in Table~\ref{tab:example_table}, we extracted a total of 58 distinct magnetic field lines spanning a wide range of magnetic activity phases, and performed numerical simulations for each case both with and without chromospheric turbulence, resulting in 116 simulation runs in total. The distributions of the coronal magnetic field strength and the coronal flux-tube expansion factor for the simulated field lines are shown in Figure~\ref{fig:extrapolated_magnetic_field_statistics}. We find a positive correlation between the coronal magnetic field strength and the coronal expansion factor \citep{Wang_2009_ApJ, Dakeyo_2024_AandA}. Since the coronal expansion factor corresponds to the local filling factor of open magnetic field regions in the corona, this positive correlation implies that regions with stronger coronal magnetic fields are predominantly dominated by closed magnetic structures, with open fields occupying only a small fraction.

\begin{figure*}
    \centering
    \includegraphics[width=0.70\linewidth]{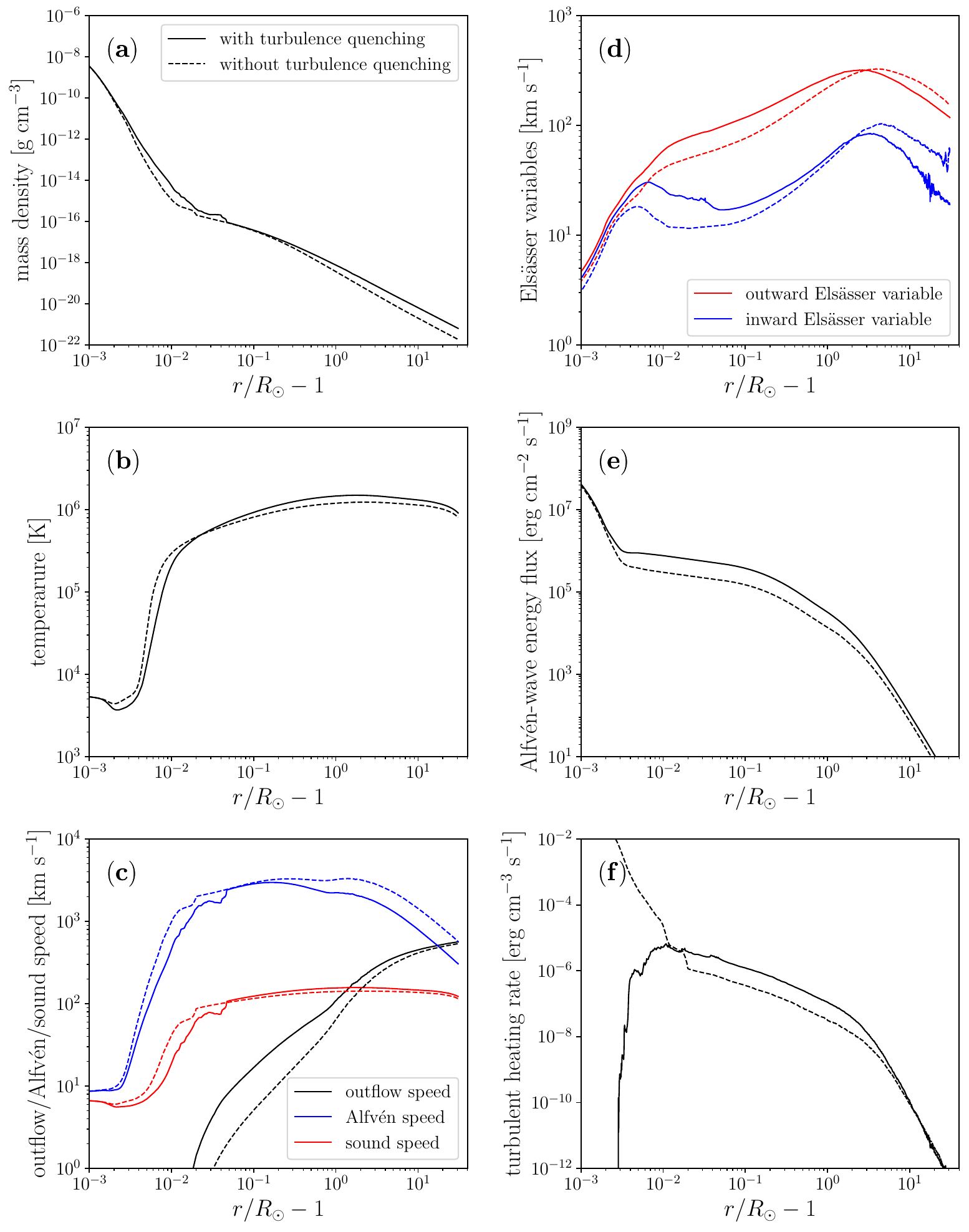}
    \caption{Radial profiles of time-averaged physical quantities with chromospheric turbulence suppression (solid lines) and without suppression (dashed lines). Panels (a) and (b) show the mass density $\langle \rho \rangle$ and temperature $\langle T \rangle$, respectively. Panel (c) displays the radial outflow velocity $\langle v_r \rangle$ (black lines) together with the Alfv\'en speed (blue lines) and sound speed (red lines). Panels (d)--(f) present the Els\"asser variables (blue: outward, red: inward), the Alfv\'en-wave energy flux, and the turbulent heating rate per unit volume.}
    \label{fig:typical_case_comparison_cr1975_lon168_sinlat52}
\end{figure*}

\subsection{Boundary condition and numerical solver}

The boundary conditions are specified as follows. Hereafter, we refer to the values of $X$ at the inner and outer boundaries as $X_{\rm in}$ and $X_{\rm out}$, respectively.

For all cases, the outer boundary is fixed at $r_{\rm out}/R_\odot = 31.8$, which typically lies in the super-Alfv\'enic region. As such, the details of the boundary condition do not significantly affect the solution, and therefore, we adopt the following boundary condition, which empirically ensures the numerical stability.
\begin{align}
    \left. \frac{\partial}{\partial r} \left( \rho r^2 \right) \right|_{\rm out} &= 0,
\end{align}
\begin{align}
    \left. \frac{\partial}{\partial r} \boldsymbol{v} \right|_{\rm out} &= 0,
\end{align}
\begin{align}
    \left. \frac{\partial}{\partial r} \left( e_{\rm int} r^3 \right) \right|_{\rm out} &= 0,
\end{align}
\begin{align}
    \left. \frac{\partial}{\partial r} \left( \boldsymbol{B}_\perp r \right) \right|_{\rm out} &= 0.
\end{align}
In a few cases without turbulence suppression and with strong coronal magnetic fields, the outer boundary falls within the sub-Alfv\'enic region, where the above reasoning may not hold. Nevertheless, since the boundary remains sufficiently distant from the subsonic region that determines the mass-loss rate, we choose to apply the same condition even in these cases.

In this study, we drive the solar wind by injecting incompressible disturbances (Alfv\'en waves) from the inner boundary. To maintain incompressibility, we impose fixed boundary conditions on the density, temperature, and radial velocity as  
\begin{equation}
    \rho_{\rm in} = 1.87 \times 10^{-7}~{\rm g~cm^{-3}}, \quad
    T_{\rm in} = 5.77 \times 10^3~{\rm K}, \quad
    v_{r,{\rm in}} = 0.
\end{equation}
The boundary conditions for the transverse components of the velocity and magnetic field are specified in terms of the Els\"asser variables defined in Equation~\eqref{eq:elsasser_definition}. For the inward-propagating Els\"asser variables, we apply a free boundary condition that allows Alfv\'en waves reflected back toward the photosphere to be transmitted out of the simulation domain:
\begin{align}
    \left. \frac{\partial}{\partial r} z_{x,y}^- \right|_{\rm in}  = 0.
\end{align}
The outward Els\"asser variables at the inner boundary are prescribed to have a broadband frequency spectrum ranging from $10^{-3}~{\rm Hz}$ to $10^{-2}~{\rm Hz}$ with a pink-noise power spectrum. The amplitude is adjusted such that the net Poynting flux at the photosphere is fixed, specifically given by
\begin{align}
    \overline{\frac{1}{4} \rho_{\rm in} \left( {z^+_{\perp,{\rm in}}}^2 - {z^-_{\perp,{\rm in}}}^2 \right) v_{\rm A, in}} 
    = 4 \times 10^8~{\rm erg~cm^{-2}~s^{-1}}, \label{eq:boundary_net_energy_flux}
\end{align}
where $v_{\rm A, in} = B_{r,{\rm in}}/\sqrt{4 \pi \rho_{\rm in}}$ is the Alfv\'en speed at the inner boundary, and the overline denotes the time average over 2000~s. This value is consistent with that adopted in a previous solar wind model \citep{Cranmer_2005_ApJ}.

We solve the basic equations by transforming all variables into cross-section-weighted forms \citep{Shoda_2021_AA}, such that the governing equations become algebraically equivalent to those in a one-dimensional Cartesian geometry. This formulation enables the use of a conventional shock-capturing MHD solver while consistently incorporating the geometrical expansion of the flux tube. The numerical fluxes are computed using the HLLD approximate Riemann solver \citep{Miyoshi_2005_JCP}.

A non-uniform spatial grid is employed. From the photosphere up to 0.03~$R_\odot$, we adopt a uniform grid spacing of 20~km. Above this height, the grid spacing gradually increases by 0.1\,\% per cell until it reaches 2000~km, beyond which it remains constant \citep{Shoda_2023_ApJ}. Spatial derivatives are evaluated using a fifth-order reconstruction scheme, achieving fifth-order spatial accuracy \citep{Suresh_1997_JCP, Mignone_2010_JCoPh, Matsumoto_2019_PASJ}, while time integration is performed with a third-order strong-stability-preserving Runge–Kutta (SSP-RK) scheme \citep{Shu_1988_JCP, Gottlieb_2001_SIAMR}. To prevent the artificial reduction of coronal density caused by under-resolved transition regions, we employ the LTRAC method \citep{Iijima_2021_ApJ}, in which the thermal-conduction coefficient and radiative-loss rate are adaptively adjusted according to the local resolution. The thermal-conduction term is advanced using a super-time-stepping technique \citep{Meyer_2012_MNRAS, Meyer_2014_JCP} to accelerate the integration of diffusive processes.

\section{Result}

\subsection{Effect of chromospheric turbulence: typical case}

To assess how the suppression of chromospheric turbulence affects the solar wind properties, we first compare representative cases with and without suppression. We specifically compare the two runs in which the background flux tube has a coronal field strength of 13.7 G and a coronal expansion factor of 4.99.

The left panels of Figure~\ref{fig:typical_case_comparison_cr1975_lon168_sinlat52} show the radial profiles of the time-averaged quantities: mass density $\left< \rho \right>$, temperature $\left< T \right>$, and radial velocity $\left< v_r \right>$, where $\left< X \right>$ denotes the time average of $X$. Panel (c) also includes the Alfv\'en speed ($B_r / \sqrt{4 \pi \langle \rho \rangle}$, blue lines) and the sound speed ($\sqrt{\langle p \rangle / \langle \rho \rangle}$, red lines) in addition to $\langle v_r \rangle$. The sound speed is evaluated under the isothermal assumption, justified by the high efficiency of thermal conduction in the corona and solar wind. Solid and dashed lines indicate the cases with and without the chromospheric turbulence suppression, respectively. Suppression of the chromospheric turbulence leads to a threefold increase in the wind density and a slight increase in the wind speed. Consequently, the mass-loss rate rises to $3.53 \times 10^{-14} M_\odot {\rm \ yr}^{-1}$, nearly 3.6 times that without suppression ($0.98 \times 10^{-14} M_\odot {\rm \ yr}^{-1}$). The increase in mass density results from a modest rise in coronal temperature, which becomes 1.2 times higher when chromospheric turbulence is suppressed. In contrast, the chromospheric temperature decreases due to reduced wave dissipation under suppressed turbulence. 

The right panels of Figure~\ref{fig:typical_case_comparison_cr1975_lon168_sinlat52} show the radial profiles of wave- and heating-related quantities, providing insight into the origin of the enhanced coronal temperature and density under chromospheric turbulence suppression. The top panel shows $\left< (z_\perp^+)^2 \right>^{1/2}$ (red lines) and $\left< (z_\perp^-)^2 \right>^{1/2}$ (blue lines), the middle panel shows $\left< F_{\rm A} \right>$ (where $F_{\rm A}$ is defined in Equation~\eqref{eq:alfven_wave_energy_density_flux}), and the bottom panel compares $\left< Q_{\rm turb} \right>$ (where $Q_{\rm turb}$ is defined in Equation~\eqref{eq:Qturb_elsasser_form}). As seen in the top panel, both the outward (blue) and inward (red) wave amplitudes at the solar surface are larger when chromospheric turbulence is suppressed. This results from reduced attenuation of the wave amplitude returning from the transition region to the photosphere, requiring a corresponding increase in the outward wave amplitude to maintain the net energy flux. Indeed, as shown in the middle panel, the energy flux at the photosphere is identical regardless of turbulence suppression. 

The middle panel shows that the energy flux decreases substantially in the chromosphere, and this decrease depends on whether turbulence is suppressed. As indicated by the turbulent dissipation rate shown in the bottom panel, turbulent dissipation in the chromosphere greatly exceeds that in the corona when chromospheric turbulence is not suppressed. This is because the dissipation rate is proportional to the mass density, which is orders of magnitude higher in the chromosphere than in the corona. Suppressing chromospheric turbulence increases both the energy flux injected into the corona and the coronal turbulent heating rate by a factor of approximately three. The enhancement of turbulent heating occurs mainly in the subsonic region, with little change in the supersonic region. This explains why the suppression of chromospheric turbulence significantly alters the mass-loss rate but has little effect on the solar wind speed in the example shown in Figure~\ref{fig:typical_case_comparison_cr1975_lon168_sinlat52}.

The results of the case study are summarised as follows. Suppressing chromospheric turbulence has a significant impact on the solar wind mass-loss rate. In particular, for the open flux tube with a coronal magnetic field of 13.7 G and a coronal expansion factor of 4.99 examined in this case study, the mass-loss rate differs by a factor of about 3.6 depending on whether chromospheric turbulence is suppressed. This is because the Alfv\'en-wave energy flux transmitted to the corona increases, leading to an enhanced turbulent heating rate in the corona. In contrast, the turbulent heating rate in the supersonic region does not significantly increase (at least in this case), and thus the solar wind speed shows little change. Based on these results, the next section discusses the outcomes of a parameter survey for various open flux tubes.

\subsection{Scaling law of mass flux}

\begin{figure}
    \centering
    \includegraphics[width=0.85\columnwidth]{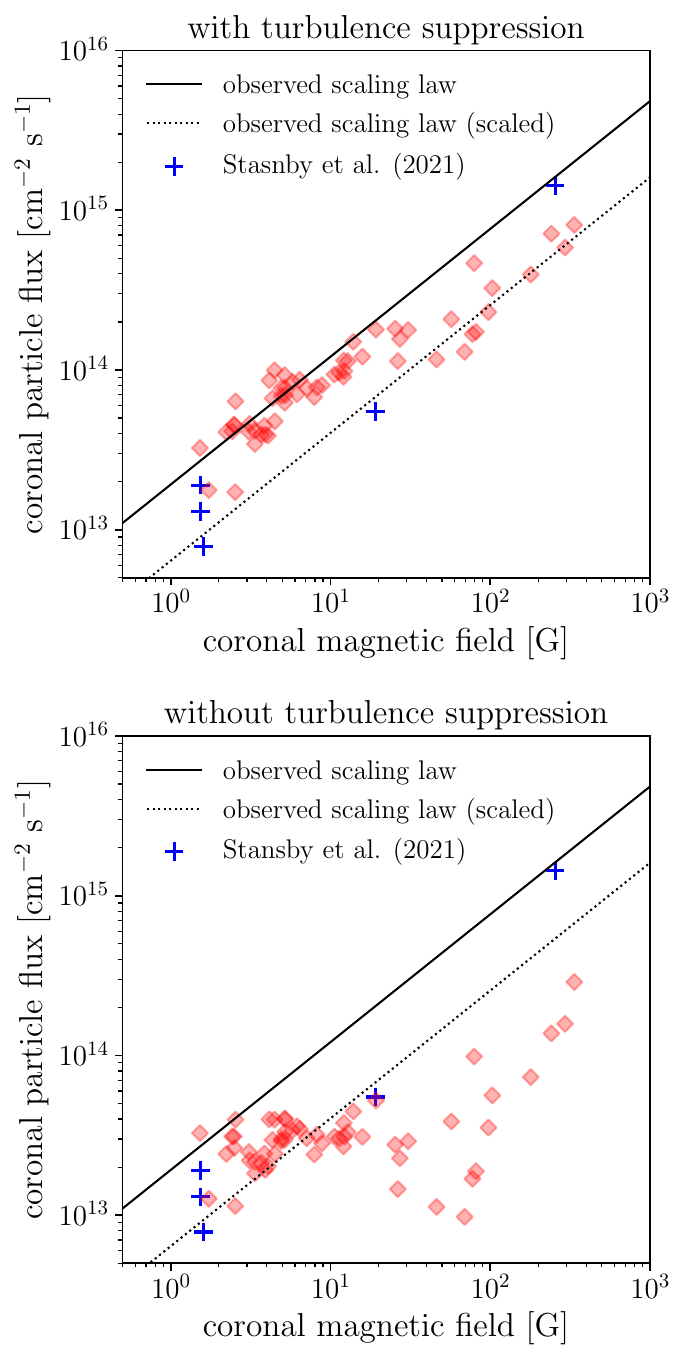}
    \caption{Coronal magnetic field--coronal particle flux relation obtained from the simulations (red diamonds), compared with the observational scaling relation of \citet{Wang_2020_ApJ} (black solid line) and its half-scaled version (black dashed line). Observational data from \citet{Stansby_2021_AandA} are also shown as blue plus symbols. Top: case with chromospheric turbulence suppression. Bottom: case without suppression.}
    \label{fig:comparison_with_Wang2020_scaling}
\end{figure}

Building on the discussion in the previous section, this section presents the results of a parameter survey of flux tubes with various coronal magnetic field strengths and coronal expansion factors (as shown in Figure~\ref{fig:extrapolated_magnetic_field_statistics}). In particular, we assess the validity of chromospheric turbulence suppression by comparing our results with the observed coronal magnetic field–proton flux scaling relation of the solar wind.

Figure~\ref{fig:comparison_with_Wang2020_scaling} shows the relationship between the coronal magnetic field ($B_{r,{\rm cb}}$) and the coronal particle flux for cases with (top) and without (bottom) chromospheric turbulence suppression. The black solid line represents the observational scaling relation derived by \citet{Wang_2020_ApJ}. The same scaling relation reduced by a factor of two is also shown as a black dashed line. Observational data from \citet{Stansby_2021_AandA} are indicated by blue plus symbols; in that study, an analysis similar to that of \citet{Wang_2020_ApJ} was carried out using different observational instruments. Red diamonds denote the coronal proton flux ($\mathcal{F}_{\rm p,cb}$), which is calculated following the procedure of \citet{Wang_2020_ApJ}:
\begin{align}
    \mathcal{F}_{\rm p,cb} = \frac{\left< \rho v_r r^2 f_{\rm cor}(r) \right>}{m_{\rm p} R_\odot^2},
\end{align}
where $m_{\rm p}$ is the proton mass. From the mass conservation, the time-averaged quantity $\left< \rho v_r r^2 f_{\rm cor}(r) \right>$ should, in principle, be independent of $r$ when averaged over a sufficiently long time interval. In practice, however, the mass flux is not spatially uniform in the chromosphere, where the shock waves dominate the radial flow. Conversely, near the outer boundary, the residual effect of the initial conditions may persist and introduce small systematic deviations. To avoid such erroneous measurements of the mass flux, we evaluate $\left< \rho v_r r^2 f_{\rm cor}(r) \right>$ at $r = 1.1 R_\odot$.

As shown in the top panel of Figure~\ref{fig:comparison_with_Wang2020_scaling}, when chromospheric turbulence is suppressed, the simulation results tend to deviate slightly from the observational scaling law near a coronal magnetic field strength of 100 G, but overall agree with the scaling derived from observations. In contrast, the bottom panel of Figure~\ref{fig:comparison_with_Wang2020_scaling} demonstrates that, without suppressing chromospheric turbulence, the simulations deviate significantly from the observational scaling relation, particularly in the strong-field regime. This suggests that suppression of chromospheric turbulence is required to reproduce a realistic mass flux within the framework of Alfv\'en wave--driven models.

In contrast to the comparison with the empirical scaling relation, a direct comparison with the individual observational data points reported by \citet{Stansby_2021_AandA} yields a more nuanced picture. For coronal magnetic fields below 20~G, the simulations without chromospheric turbulence suppression are in slightly better agreement with the observations. By contrast, at the high magnetic field strength of 255~G reported in the same study, the model including chromospheric turbulence suppression shows better consistency. This behaviour suggests that outflows originating from strong-field regions may provide a useful means of distinguishing between the two modelling approaches. Given that the magnetic field strength in solar coronal holes is typically in the range of 5--10~G, models intended to represent real coronal holes are unlikely to exhibit a substantial difference between cases with and without chromospheric turbulence suppression.

\begin{figure}
    \centering
    \includegraphics[width=0.95\columnwidth]{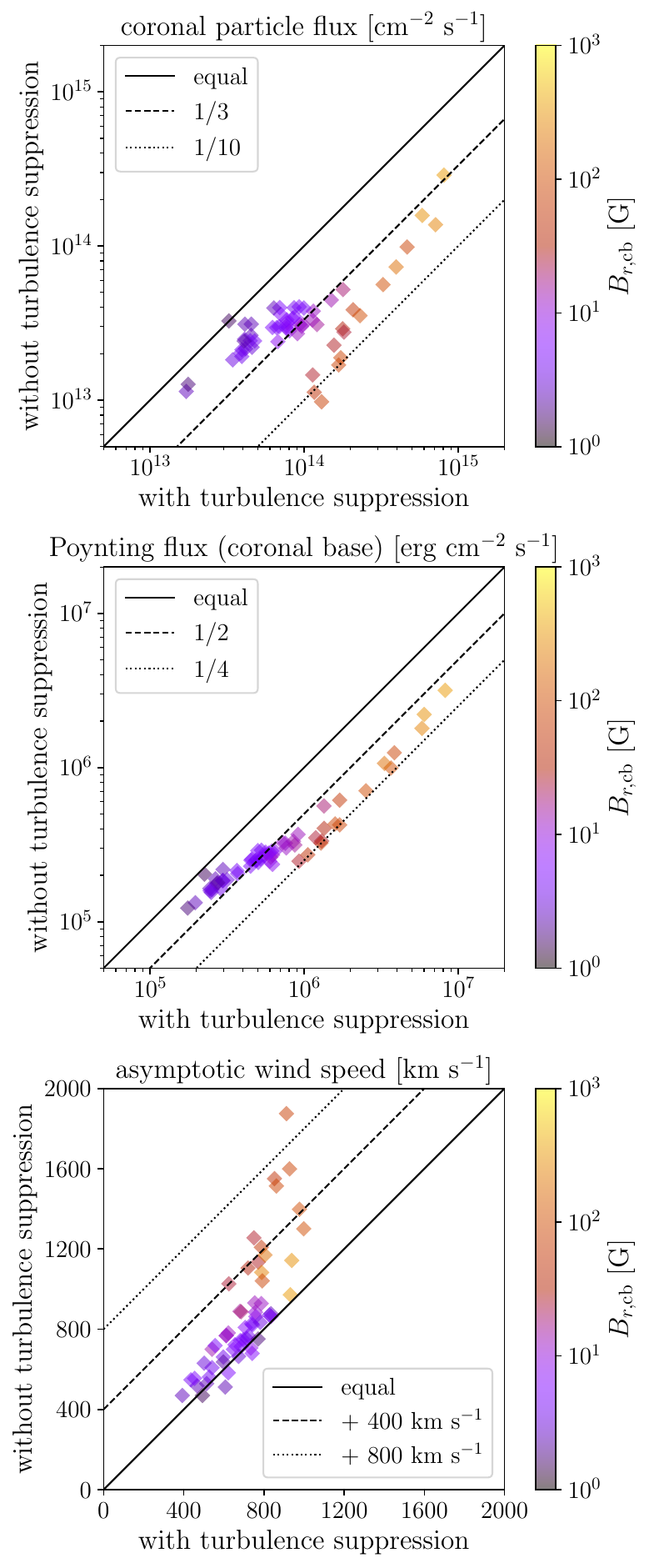}
    \caption{Comparison of simulation outputs obtained with and without chromospheric turbulence suppression. The horizontal axis represents the values from the simulations with turbulence suppression, and the vertical axis represents those from the simulations without suppression. The top, middle, and bottom panels show the coronal particle flux, the Poynting flux at the coronal base, and the asymptotic wind speed, respectively. The colour of each symbol indicates the magnetic field strength at the coronal base. The black solid line indicates the one-to-one correspondence between the two cases, while the dashed and dotted lines represent scaled versions of the solid line for reference.}
    \label{fig:comparison_with_without_suppression}
\end{figure}

To investigate why the mass flux differs by up to an order of magnitude depending on whether chromospheric turbulence is suppressed, we examined the global energy budget of the stellar wind. From the energy conservation law, the coronal proton flux can be approximately expressed in terms of the Poynting flux at the coronal base, $F_{\rm A,cb}$, and the asymptotic wind speed, $v_\infty$, as follows \citep{Cranmer_2011_ApJ, Shoda_2020_ApJ}:
\begin{align}
    \mathcal{F}_{\rm p,cb} \approx \frac{F_{\rm A, cb}}{m_{\rm p} \left( v_{\rm esc}^2 + v_\infty^2 \right)/2},
\end{align}
where $v_{\rm esc} = 617 {\rm \ km \ s^{-1}}$ denotes the escape velocity of the Sun. We therefore examine how these two quantities respond to chromospheric turbulence suppression to identify the primary cause of the large variation in the modeled solar-wind mass flux.

\begin{figure}
    \centering
    \includegraphics[width=0.95\columnwidth]{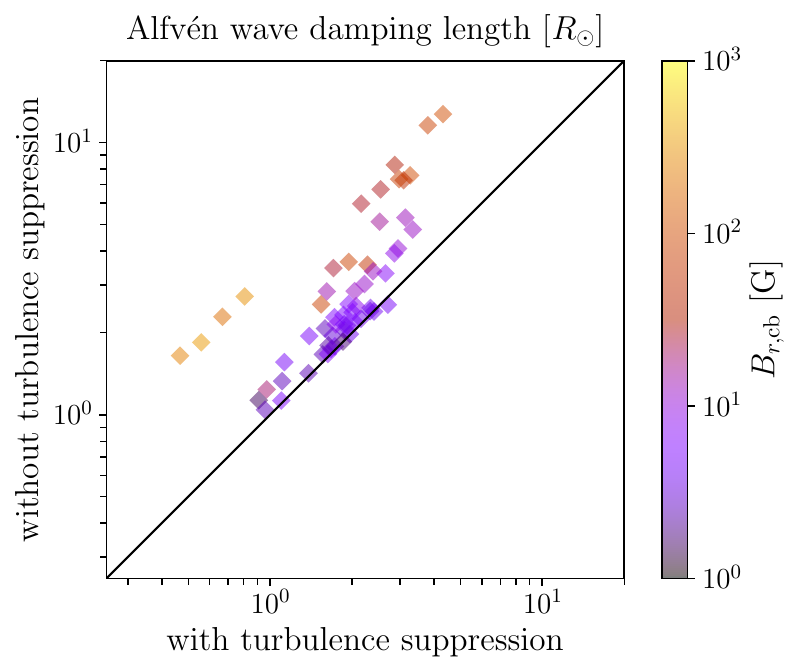}
    \caption{Comparison of the Alfv\'en-wave damping length $L_{\mathrm{D}}$ (defined in Eq.~\eqref{eq:wave_damping_length}) between models with and without chromospheric turbulence suppression. The horizontal axis shows the values obtained with suppression, and the vertical axis those without suppression. The colour of each symbol denotes the magnetic-field strength at the coronal base. The black solid line marks the one-to-one relation.}
    \label{fig:Alfven_wave_damping_length}
\end{figure}


Figure~\ref{fig:comparison_with_without_suppression} compares the coronal particle flux (top panel), coronal Poynting flux $F_{\rm A,cb}$ (middle panel), and asymptotic wind speed (bottom panel) between cases with and without chromospheric turbulence suppression, in the form of scatter plots. Each symbol corresponds to the results from the two models for an individual magnetic field line. As seen in the top panel, the coronal particle flux systematically increases when chromospheric turbulence is suppressed, and the magnitude of this increase tends to be larger for stronger coronal magnetic fields. Specifically, the difference in coronal particle flux between the two models exceeds an order of magnitude in two cases. It should be noted, however, that these two cases correspond to moderately strong coronal magnetic fields of $B_{r,\rm{cb}} = 69$~G and $B_{r,\rm{cb}} = 46$~G, rather than to the most extreme cases. For the strongest field case in our simulations ($B_{r,\rm{cb}} = 336$~G), the ratio is limited to 2.81.

The middle panel of Figure~\ref{fig:comparison_with_without_suppression} shows a scatter plot of the coronal-base Poynting flux. For $B_{r,\mathrm{cb}} \leq 20$~G, the distribution exhibits a trend similar to that of the coronal particle flux. However, for $B_{r,\mathrm{cb}} \geq 20$~G, the deviation between the cases with and without chromospheric turbulence suppression is limited to a factor of $\sim 4$, which is smaller than the difference found for the coronal particle flux. This indicates that the gap in coronal particle flux between the two models can be explained by the difference in Poynting flux for $B_{r,\mathrm{cb}} \leq 20$~G, but not for $B_{r,\mathrm{cb}} \geq 20$~G. Indeed, as shown in the bottom panel of Figure~\ref{fig:comparison_with_without_suppression}, the asymptotic wind speeds are almost identical between the two models for $B_{r,\mathrm{cb}} \leq 20$~G, whereas for $B_{r,\mathrm{cb}} \geq 20$~G the wind speed is significantly larger in the case without chromospheric turbulence suppression. The difference reaches up to $\sim 1000~\mathrm{km~s^{-1}}$, with $v_\infty = 916~\mathrm{km~s^{-1}}$ in the case with chromospheric turbulence suppression and $v_\infty = 1870~\mathrm{km~s^{-1}}$ in the case without suppression. This difference in asymptotic wind speed accounts for a factor of $3.17$ difference in the coronal particle flux between the two models.


The variation in the solar wind speed appears to arise from a change in the location of Alfv\'en-wave dissipation associated with the presence or absence of chromospheric turbulence. As shown in Figure~\ref{fig:typical_case_comparison_cr1975_lon168_sinlat52}e, suppressing chromospheric turbulence markedly enhances the heating rate in the subsonic region, while leaving it nearly unchanged in the supersonic region. This indicates that turbulence suppression shifts the dissipation preferentially towards lower altitudes, thereby modifying the distribution of energy deposition relative to the sonic point.

To quantify this effect, we analyse the effective damping length of Alfv\'en waves in the corona. The damping length, $L_{\rm D}$, is defined from an exponential approximation to the radial profile of the Poynting flux as follows
\begin{align}
    \langle F_{\rm A} r^2 f \rangle \propto \exp \left(-\frac{r - r_{\rm cb}}{L_{\rm D}}\right), \label{eq:wave_damping_length}
\end{align}
where angle brackets denote a time average and the subscript cb refers to the coronal base. Since our primary interest lies in the subsonic corona, we perform the fitting over the radial range $1.01 < r/R_\odot < 2$ and determine $L_{\rm D}$ for each case.

Figure~\ref{fig:Alfven_wave_damping_length} shows a scatter plot comparing $L_{\rm D}$ between simulations with and without chromospheric turbulence suppression. The colour of the symbols represents the coronal magnetic field strength. For relatively weak coronal magnetic fields, no significant difference in the damping length is observed between the two cases. However, as the coronal magnetic field increases, the simulations with chromospheric turbulence suppression exhibit systematically shorter damping lengths. This indicates that suppressing chromospheric turbulence enhances the dissipation of Alfv\'en-wave energy closer to the Sun, which in turn contributes to the reduction of the terminal solar wind speed \citep{Leer_1980_JGR, Hansteen_1995_JGR}.


\section{Discussion}

\subsection{Sensitivity to the chromospheric turbulence: physical origin \label{sec:toy_model}}

One of the key objectives of this work is to quantify the sensitivity of the solar-wind mass flux to chromospheric turbulence dissipation. Our simulations show that suppressing chromospheric turbulence leads to an increase in mass flux by up to an order of magnitude, depending on the coronal magnetic field strength. This result demonstrates that chromospheric dissipation has a significant impact on the mass loading of the solar wind.

\begin{figure}
    \centering
    \includegraphics[width=0.85\columnwidth]{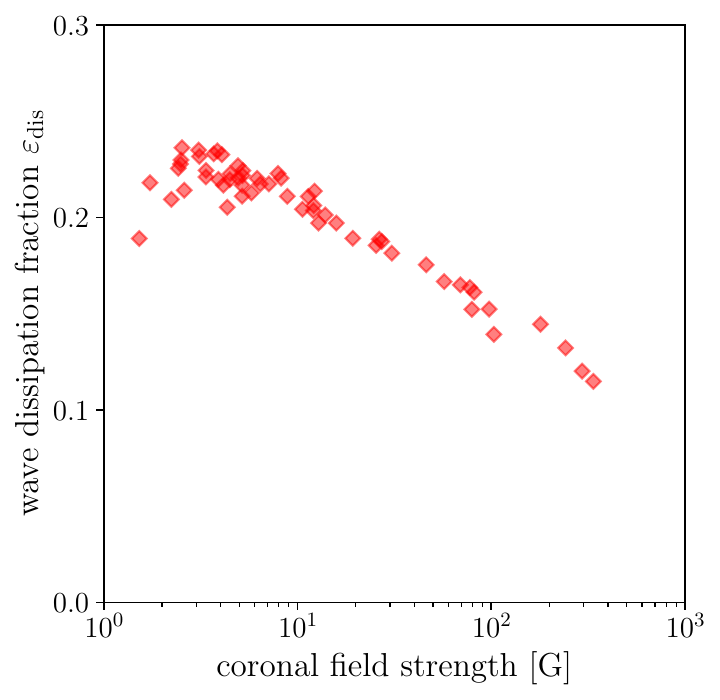}
    \caption{Normalized chromospheric turbulent dissipation rate of Alfv\'en wave (vertical axis, defined in Eq.~\eqref{eq:epsilon_dis}), as a function of the coronal magnetic-field strength (horizontal axis).}
    \label{fig:turbulence_timescale_decayrate}
\end{figure}

It is important to note that our simulation results are not entirely trivial. The Alfv\'en crossing time in the chromosphere, estimated from a typical Alfv\'en speed of 10--100 km s$^{-1}$ and a transition-region height of 2--10 Mm, is on the order of a few hundred seconds. In contrast, the turbulent dissipation timescale of upward-propagating Alfv\'en waves in the chromosphere can be estimated as
\begin{align}
    \tau_{\rm dis} = \frac{\lambda_\perp}{c_{\rm dis} z^-_\perp}
    \approx \frac{\lambda_{\perp,\odot}}{c_{\rm dis} z^-_{\perp,\odot}}
    \sim 10^{3}\ {\rm s},
\end{align}
which is longer than the propagation time. This ordering implies that a substantial fraction of the upward Alfv\'en-wave energy can traverse the chromosphere without being dissipated locally.

To quantify the relative importance of chromospheric Alfv\'en-wave dissipation and its dependence on magnetic-field strength, we define a normalized dissipation rate,
\begin{equation}
    \epsilon_{\rm dis} = \int_{r_{\rm in}}^{r_{\rm tr}} \frac{dr}{\tau_{\rm dis} v_{\rm A}}, \label{eq:epsilon_dis}
\end{equation}
where $r_{\rm tr}$ denotes the radial distance of the transition region and $v_{\rm A} = B_r / \sqrt{4 \pi \langle \rho \rangle}$. Here, $r_{\rm tr}$ is defined as the height at which the time-averaged temperature first exceeds $10^5$~K. The quantity $\epsilon_{\rm dis}$ represents the fractional energy loss experienced by Alfv\'en waves as they propagate from the photosphere (the inner boundary) to the transition region.

Figure~\ref{fig:turbulence_timescale_decayrate} shows $\epsilon_{\rm dis}$ as a function of the coronal magnetic-field strength for models without chromospheric turbulence suppression, as $\epsilon_{\rm dis}$ is negligible when suppression is applied. The normalized dissipation rate typically lies in the range $\epsilon_{\rm dis} \simeq 0.1$--0.25 and decreases with increasing magnetic-field strength, indicating that chromospheric turbulent dissipation becomes less efficient in stronger magnetic fields. Nevertheless, our simulations demonstrate that chromospheric turbulence still has a significant influence on the solar-wind mass flux, and that this influence becomes stronger as the coronal magnetic field increases.

\begin{figure}
    \centering
    \includegraphics[width=1.0\columnwidth]{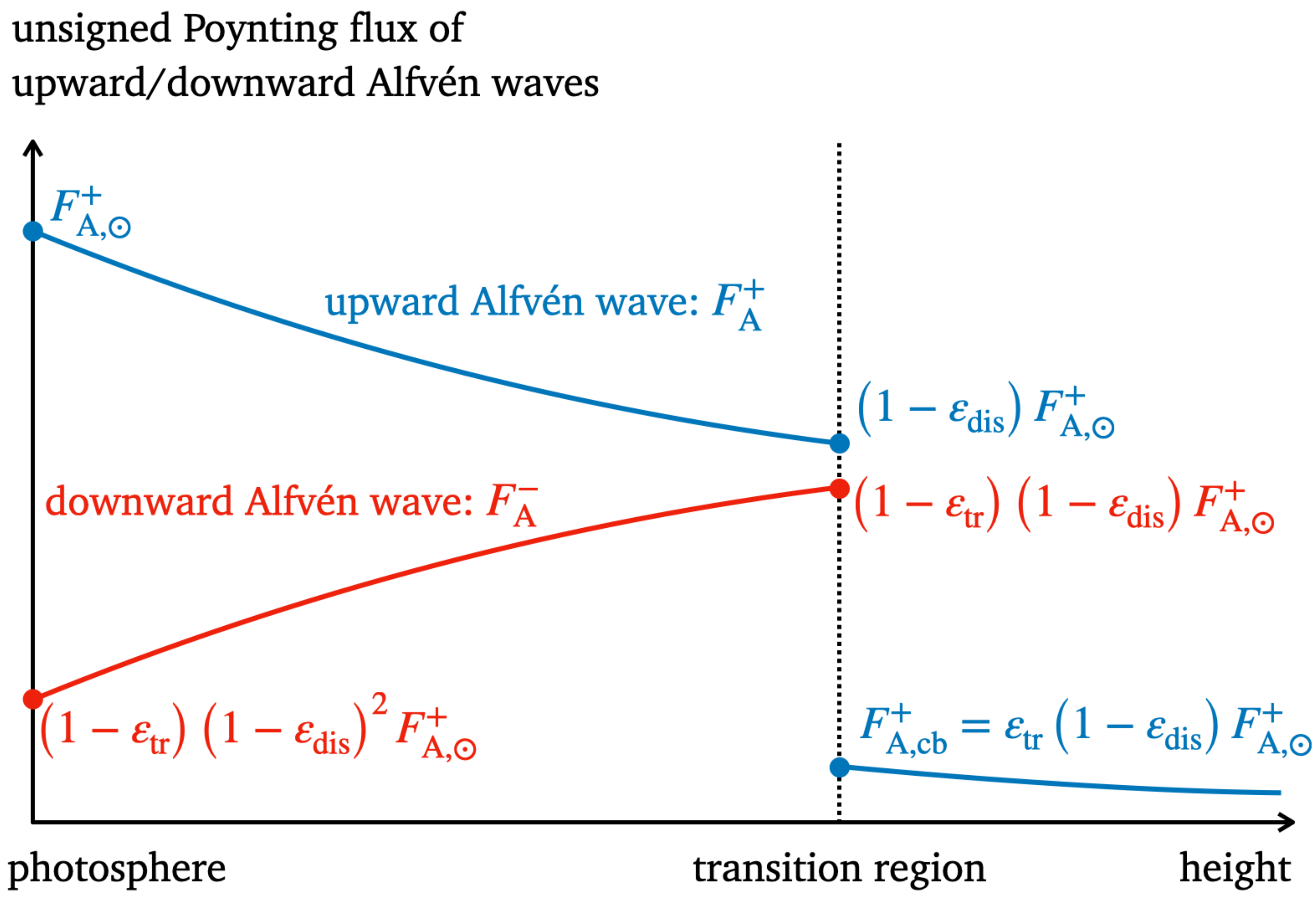}
    \vspace{-1em}
    \caption{Schematic illustration of the toy model for energy transport from the photosphere to the corona. The horizontal axis represents the height above the stellar surface, and the vertical axis shows the unsigned Poynting flux. The vertical dashed line indicates the location of the transition region. The blue line shows the height profile of the Poynting flux carried by upward-propagating Alfv\'en waves ($F_{\rm A}^+$), and the red line shows that carried by downward-propagating Alfv\'en waves ($F_{\rm A}^-$).
    \label{fig:energyflux_toymodel}}
\end{figure}

This counterintuitive behavior can be interpreted as the result of the interplay between wave dissipation in the chromosphere and reflection in the upper chromosphere and the transition region. To clarify this effect in a simplified form, we introduce the toy model described in Figure~\ref{fig:energyflux_toymodel}. We assume that Alfv\'en waves lose a fraction $\varepsilon_{\rm dis}$ of their energy flux while propagating through the chromosphere, and that a fraction $\varepsilon_{\rm tr}$ of the remaining flux is transmitted through the transition region (so that $1-\varepsilon_{\rm tr}$ is reflected). We then treat all reflections occurring between the upper chromosphere and the transition region as being effectively assigned to the transition region, so that the model assumes a single reflection layer located there. Let $F_{{\rm A},\odot}^{+}$ denote the upward Alfv\'en-wave energy flux at the photosphere. Based on the assumptions of the toy model, the upward energy flux immediately below the transition region is $\left( 1-\varepsilon_{\rm dis} \right) F_{{\rm A},\odot}^{+}$, and the corresponding downward flux is $\left( 1 - \varepsilon_{\rm tr} \right)\left( 1- \varepsilon_{\rm dis} \right) F_{{\rm A},\odot}^{+}$. Using these relations, the downward energy flux at the photosphere can be written as 
\begin{equation}
    F_{{\rm A},\odot}^- = \left( 1 - \varepsilon_{\rm tr} \right)\left( 1- \varepsilon_{\rm dis} \right)^2 F_{{\rm A},\odot}^{+}.
\end{equation}
The net Poynting flux at the photosphere is therefore
\begin{align}
  F_{{\rm A}, \odot}
  &= F_{{\rm A},\odot}^{+} - F_{{\rm A},\odot}^{-} \notag \\
  &= F_{{\rm A},\odot}^{+} - \left( 1 - \varepsilon_{\rm tr} \right)\left( 1- \varepsilon_{\rm dis} \right)^2 F_{{\rm A},\odot}^{+} \notag \\
  &= \big( \varepsilon_{\rm tr} + 2\varepsilon_{\rm dis} - \varepsilon_{\rm dis}^{2}
- 2\varepsilon_{\rm tr}\varepsilon_{\rm dis} + \varepsilon_{\rm tr}\varepsilon_{\rm dis}^{2} \big) F_{{\rm A},\odot}^{+}.
\end{align}
For simplicity, we assume that no downward-propagating Alfv\'en waves are present in the corona. The net energy flux at the coronal base then reduces to
\begin{align}
    F_{\mathrm{A,cb}} = F_{\mathrm{A,cb}}^{+} = \varepsilon_{\rm tr} \left( 1 - \varepsilon_{\rm dis} \right) F_{{\rm A},\odot}^{+}.
\end{align}
Under this assumption, the net energy transmission rate between the photosphere and the coronal base (denoted as TR) can be written as the ratio of the net energy fluxes at these two locations:
\begin{align}
    {\rm TR} = \frac{F_{{\rm A},{\rm cb}}}{F_{{\rm A},\odot}} = \frac{\varepsilon_{\rm tr} \left( 1 - \varepsilon_{\rm dis} \right)}{\varepsilon_{\rm tr} + 2\varepsilon_{\rm dis} - \varepsilon_{\rm dis}^{2}
- 2\varepsilon_{\rm tr}\varepsilon_{\rm dis} + \varepsilon_{\rm tr}\varepsilon_{\rm dis}^{2}}. \label{eq:net_transmission_rate}
\end{align}

Equation~\eqref{eq:net_transmission_rate} shows that the net transmission of wave energy from the photosphere to the corona is controlled by the relative importance of chromospheric dissipation and transition-region reflection. To clarify its limiting behaviour, we consider two asymptotic regimes. In the limit $\varepsilon_{\rm dis} \ll \varepsilon_{\rm tr} < 1$, the transmission rate can be approximated as
\begin{align}
    {\rm TR} \approx \frac{\varepsilon_{\rm tr} \left( 1-\varepsilon_{\rm dis} \right)}{\varepsilon_{\rm tr} + 2\varepsilon_{\rm dis} - 2\varepsilon_{\rm tr}\varepsilon_{\rm dis}} \approx 1 - \frac{\varepsilon_{\rm dis}}{\varepsilon_{\rm tr}} \left( 2 - \varepsilon_{\rm tr} \right) \approx 1,
\end{align}
which indicates that, when chromospheric dissipation is sufficiently weak, nearly all of the net energy flux is transmitted from the photosphere to the corona, regardless of the transmission coefficient of the transition region. When the transmission coefficient of the transition region becomes small, the upward energy flux reaching the corona decreases. At the same time, the reflected component increases correspondingly and propagates back towards the photosphere, thereby reducing the net Poynting flux injected from below. Because the transmission rate is defined as the ratio of net fluxes, these two effects compensate each other, so that ${\rm TR} \approx 1$ as long as $\varepsilon_{\rm dis} \ll \varepsilon_{\rm tr}$.

Conversely, in the opposite regime ($\varepsilon_{\rm tr} \ll \varepsilon_{\rm dis} < 1$), the transmission rate reduces to
\begin{align}
    {\rm TR} \approx \frac{\varepsilon_{\rm tr}}{\varepsilon_{\rm dis}} \frac{1 - \varepsilon_{\rm dis}}{2 - \varepsilon_{\rm dis}} \sim \frac{\varepsilon_{\rm tr}}{\varepsilon_{\rm dis}},
\end{align}
where we have used the fact that $\left( 1-\varepsilon_{\rm dis}\right)/\left(2-\varepsilon_{\rm dis}\right)$ remains of order unity for $\varepsilon_{\rm dis}<1$. This result indicates that even weak chromospheric dissipation ($\varepsilon_{\rm dis} \ll 1$) leads to a net transmission rate much smaller than unity if the transition region is highly reflective ($\varepsilon_{\rm tr} \ll \varepsilon_{\rm dis}$).

On the basis of these results, a physical interpretation can be formulated. Because the turbulent correlation length at the photosphere is prescribed identically in all cases and the amplitude of the incident Alfv\'en waves remains nearly constant, the characteristic timescale of turbulent dissipation exhibits little dependence on the coronal magnetic-field strength. By contrast, an increase in the coronal magnetic field leads to a higher Alfv\'en speed in the upper chromosphere and corona, which in turn enhances the wave-reflection coefficient near the transition region. As a consequence, the reduction of the energy-transmission efficiency associated with a decrease in $\varepsilon_{\rm tr}$ becomes more pronounced for stronger fields, thereby amplifying the impact of chromospheric turbulence on the solar-wind mass flux.

\subsection{On the origin of turbulence suppression in the chromosphere}

In this study, we explored the possibility that chromospheric turbulence is weaker than commonly assumed, motivated by an alternative picture of wave excitation. Specifically, we consider a scenario in which Alfv\'enic perturbations are driven by coherent rotational motions of flux tubes \citep{Fedun_2011_AnGeo, Morton_2013_ApJ, Skirvin_2025_ApJ}. Such waves have attracted increasing attention in the context of energy transport and heating of the upper solar atmosphere \citep{Kuniyoshi_2024_ApJ, Kuniyoshi_2025_ApJ}. It differs from conventional models that attribute the excitation primarily to sub-scale flows within photospheric magnetic elements \citep{van_Ballegooijen_2011_ApJ, van_Ballegooijen_2017_ApJ_coronal_loop}. However, high-resolution radiative MHD simulations suggest that vortex-driven Alfv\'en waves may also be excited within single flux tubes \citep{Yadav_2020_ApJ, Yadav_2021_AandA}. In this case, turbulence can still develop within the tube, and chromospheric turbulent dissipation may remain comparable to that in conventional models.

In addition to the excitation scenario discussed above, the nature of the wave modes themselves may also act to suppress chromospheric turbulence. Photospheric magnetic patches are known to move horizontally due to surface thermal convection \citep{Berger_1996_ApJ, Utz_2010_AandA}. In this situation, the Poynting flux is expected to be transported mainly by the transverse displacement of entire flux tubes, that is, by kink waves \citep{Zaitsev_1975_IGAFS, Wentzel_1979_ApJ, Cranmer_2005_ApJ, Fujimura_2009_ApJ, Stangalini_2014_AandA}. When the kink mode dominates, the chromospheric turbulence is more appropriately described by the uni-turbulence model \citep{Magyar_2017_NatSR}, which is characterized by a damping rate that depends on the magnetic filling factor and vanishes when the filling factor reaches unity \citep{Van_Doorsselaere_2025_AndA}. Because the magnetic filling factor at the photosphere increases in strong-field regions \citep{Solanki_1992_AandA}, the filling factor should reach unity at lower heights in the chromosphere. As a result, most of the chromospheric volume becomes ineffective for the damping by uni-turbulence damping under strong-field conditions, leading to a substantial reduction in the overall damping rate. This effect could help resolve the apparent saturation of coronal energy input in the presence of strong magnetic fields.

A quantitative assessment of these mechanisms requires analysis beyond reduced MHD \citep{Carbone_2009_PhRvL, Yokoi_2018_JPlPh, Finley_2022_AA}. In particular, fully compressible MHD simulations combined with diagnostics of scale-to-scale energy transfer \citep{Dong_2022_SciA} and higher-order statistical measures \citep{Roberts_2022_JGRA, Wang_2024_AandA} will be necessary to characterise the development and suppression of turbulence under these excitation and modal conditions. A systematic investigation along these lines is left for future work.

\subsection{On the choice of boundary condition for wave energy flux}

As specified in Eq.~\eqref{eq:boundary_net_energy_flux}, our model adopts a boundary condition in which the net wave energy flux, rather than the incident flux, is held fixed at the lower boundary. As indicated by the toy model presented in Section~\ref{sec:toy_model}, this assumption is likely to play a key role in determining the resulting trends. However, the assumption of a fixed net energy flux is not uniquely supported by current theoretical or observational constraints. Therefore, the quantitative results presented in this study may depend on the adopted boundary condition, and caution is required when interpreting the model predictions.

The choice to fix the net energy flux is motivated as follows. The vertical Poynting flux at the photosphere can be approximated as the sum of contributions from horizontal footpoint motions (the shaking term) and flux emergence (the emergence term), while the resistive term is negligible \citep{Shelyag_2012_ApJ, Finley_2022_AA}:
\begin{align}
    S_{\rm v} \approx - \frac{B_{\rm v}}{4 \pi} \left( \boldsymbol{v}_{\rm h} \cdot \boldsymbol{B}_{\rm h} \right) 
    + \frac{B_{\rm h}^2}{4 \pi} v_{\rm v}.
\end{align}
Here, $B_{\rm v}$ is the vertical magnetic field, and $\boldsymbol{v}_{\rm h}$ and $\boldsymbol{B}_{\rm h}$are the horizontal components of velocity and magnetic field, respectively. In general, both the shaking and emergence terms can contribute comparably to the total energy transport \citep{Finley_2022_AA}.

In this study, however, we focus on the energy injection associated with Alfv\'enic wave excitation driven by horizontal convective motions. We therefore consider only the shaking component of the Poynting flux. Under this assumption, the statistically averaged net upward energy flux associated with wave driving is given by
\begin{equation}
    F_{\rm net} \approx - \frac{B_{\rm v}}{4 \pi} \langle \boldsymbol{v}_{\rm h} \cdot \boldsymbol{B}_{\rm h} \rangle,
\end{equation}
where $\langle \cdot \rangle$ denotes a statistical average. In surface convection, the correlation $\langle \boldsymbol{v}_{\rm h} \cdot \boldsymbol{B}_{\rm h} \rangle$ can exhibit a systematic bias depending on the sign of $B_{\rm v}$, resulting in a finite net energy flux. We assume that this velocity--magnetic field correlation is primarily determined by the intrinsic properties of photospheric convection, and is therefore largely insensitive to the physical processes occurring in the overlying chromosphere and transition region, such as wave reflection. Under this assumption, the net Poynting flux is effectively controlled at the photosphere, making it natural to adopt a boundary condition in which the net flux, or equivalently $\langle \boldsymbol{v}_{\rm h} \cdot \boldsymbol{B}_{\rm h} \rangle$, is fixed.

We emphasize that our treatment represents a modeling assumption rather than a definitive physical conclusion. An alternative possibility is that convection determines the incident wave flux, while the net flux is regulated through interaction with the overlying chromosphere and transition region \citep{Shoda_2018_ApJ_a_self-consistent_model, Shoda_2020_ApJ}. In this case, following the toy model presented in Section~\ref{sec:toy_model}, the net energy flux would be reduced by a factor of $\varepsilon_{\rm tr}$ (the transmission efficiency), and the mass loss rate may even decrease with increasing magnetic field strength. Discriminating between these scenarios would require self-consistent simulations that couple the convection zone with the upper atmosphere, which is left for future work.

\subsection{Other limitations and implications}

In addition to the assumption adopted for the boundary condition, several caveats should be kept in mind when interpreting the present results. The first concerns the validity of the model–observation comparison. The empirical relationship derived by \citet{Wang_2020_ApJ} was obtained by tracing the solar wind measurements at 1 au back to their coronal sources. However, the traceback process is inherently imperfect because low-latitude solar wind streams undergo mixing with surrounding flows during their propagation \citep{McGregor_2011_JGRA, Riley_2011_SoPh}. A strictly consistent comparison would require forward modeling based on global solar wind simulations \citep{Odstrcil_2003_AdSpR, Shiota_2014_SpaceWeather, Pomoell_2018_JSWSC, Perri_2022_ApJ} that naturally include mixing effects. Because the present model describes the wind evolution only within a single flux tube, such effects are not captured. In this regard, although the model is found to reproduce the empirical relation when chromospheric turbulence is suppressed, this agreement may be partly coincidental.

The second point to note is that, although suppressing chromospheric turbulence leads to a better agreement between the model and the observations in terms of mass flux, this does not necessarily imply that the specific form of turbulence suppression adopted in our model is physically realistic. The temperature-dependent prescription we employed is not based on a concrete physical mechanism, and completely turning off turbulent dissipation represents an extreme treatment. Our results therefore support the qualitative conclusion that chromospheric turbulence has a significant impact on the wind mass flux, although they do not provide quantitative constraints on how turbulence should be suppressed in reality.

We also note that ambipolar diffusion may also play a role in the chromosphere \citep{Khomenko_2012_ApJ, Martinez_Sykora_2017_Sci, Soler_2026_AA}. One might expect that enhanced diffusion could lead to stronger dissipation and potentially suppresses the wind mass flux \citep{Matsuoka_2024_ApJ, Suzuki_2025_PASJ}. However, its impact on Alfv\'en-wave turbulence is not straightforward. Ambipolar diffusion preferentially acts on currents perpendicular to the magnetic field \citep{Khomenko_2014_PhPl}, whereas Alfv\'en-wave turbulence tends to generate field-aligned currents through the perpendicular cascade process \citep{Shebalin_1983_JPP, Goldreich_1995_ApJ, Cho_2003_MNRAS, Howes_2016_ApJ}. Therefore, the selective dissipation of perpendicular currents does not necessarily imply a simple enhancement of turbulent dissipation. A quantitative assessment of this effect would require self-consistent turbulence simulations including ambipolar diffusion, which is beyond the scope of the present study.

Despite the caveats discussed above, our results provide several important implications for stellar wind modeling. First, when chromospheric turbulence is not suppressed, the mass-loss rate does not increase with magnetic activity \citep{Shoda_2020_ApJ}, implying that additional energy input mechanisms such as interchange reconnection are required to reproduce the observed scaling relations \citep{Shoda_2023_ApJ}. Our results suggest that, even without introducing such mechanisms, wave-driven models alone may produce mass-loss rates that increase with magnetic activity when chromospheric turbulence is properly accounted for. This highlights the potentially critical role of chromospheric processes in controlling stellar wind mass loss.

A further implication concerns the treatment of the lower boundary in global solar and stellar wind models. Most existing models \citep{van_der_Holst_2014_ApJ} set their lower boundary just below the transition region, in the upper chromosphere. Our findings imply that extending the lower boundary down to the stellar surface requires particular attention to the treatment of wave propagation and turbulent dissipation in the chromosphere, especially for stars with strong magnetic activity.

\section{Conclusions}

In this study, we investigate how the chromospheric turbulence affects the mass flux of solar and stellar winds using one-dimensional wave-driven wind simulations. We find that suppressing the chromospheric turbulence leads to a systematic increase in the coronal particle flux, by up to an order of magnitude, particularly in moderately strong magnetic field regions. This behavior arises from a combination of changes in the Poynting flux at the coronal base and in the asymptotic wind speed. We also show that this treatment naturally reproduces the observed empirical relation between coronal magnetic field strength and mass flux without invoking additional energy input mechanisms such as interchange reconnection. These findings highlight the critical role of the chromospheric turbulence as a key factor in the stellar wind mass flux and suggest that such effects should be carefully treated in models that connect the stellar surface and the stellar wind.

\section*{Acknowledgements}

Numerical computations were carried out on the Cray XD2000 at the Center for Computational Astrophysics, National Astronomical Observatory of Japan, and on the Wisteria/BDEC-01 Odyssey systems at the University of Tokyo. Part of the computational resources used in this study was provided by the Joint Usage/Research Center for Interdisciplinary Large-scale Information Infrastructures (JHPCN) in Japan (Project ID: jh230046). MS was supported by JSPS KAKENHI Grant Number JP24K00688. TVD was supported by a Senior Research Project (G088021N) of the FWO Vlaanderen. Furthermore, TVD received financial support from the Flemish Government under the long-term structural Methusalem funding program, project SOUL: Stellar evolution in full glory, grant METH/24/012 at KU Leuven. The research that led to these results was subsidised by the Belgian Federal Science Policy Office through the contract B2/223/P1/CLOSE-UP. It is also part of the DynaSun project and has thus received funding under the Horizon Europe programme of the European Union under grant agreement (no. 101131534). Views and opinions expressed are however those of the author(s) only and do not necessarily reflect those of the European Union and therefore the European Union cannot be held responsible for them. ASB acknowledges funding by the European Research Council via ERC Synergy grant No. 810218 (“The Whole Sun”, ERC-2018-SyG), CNES Solar Orbiter support and CNRS/INSU AT-ST grant. TVD would like to thank MS for the hospitality and discussion during his stay at Tokyo University in April 2025. ASB is also thankful to the University of Nagoya and ISEE for their support as a designated full professor and to Prof. Hideyuki Hotta.

\section*{Data Availability Statement}
The numerical code and data underlying this article will be shared on reasonable request to the corresponding author.



\bibliographystyle{mnras}
\bibliography{reference_new} 

@ARTICLE{Airapetian_2017_ApJ,
       author = {{Airapetian}, Vladimir S. and {Glocer}, Alex and {Khazanov}, George V. and {Loyd}, R.~O.~P. and {France}, Kevin and {Sojka}, Jan and {Danchi}, William C. and {Liemohn}, Michael W.},
        title = "{How Hospitable Are Space Weather Affected Habitable Zones? The Role of Ion Escape}",
      journal = {\apjl},
         year = 2017,
        month = feb,
       volume = {836},
       number = {1},
          eid = {L3},
        pages = {L3},
          doi = {10.3847/2041-8213/836/1/L3},
       adsurl = {https://ui.adsabs.harvard.edu/abs/2017ApJ...836L...3A}
}

@ARTICLE{Airapetian_2021_ApJ,
       author = {{Airapetian}, Vladimir S. and {Jin}, Meng and {L{\"u}ftinger}, Theresa and {Boro Saikia}, Sudeshna and {Kochukhov}, Oleg and {G{\"u}del}, Manuel and {Van Der Holst}, Bart and {Manchester}, W., IV},
        title = "{One Year in the Life of Young Suns: Data-constrained Corona-wind Model of {\ensuremath{\kappa}}$^{1}$ Ceti}",
      journal = {\apj},
         year = 2021,
        month = aug,
       volume = {916},
       number = {2},
          eid = {96},
        pages = {96},
          doi = {10.3847/1538-4357/ac081e},
archivePrefix = {arXiv},
       eprint = {2106.01284},
 primaryClass = {astro-ph.SR},
       adsurl = {https://ui.adsabs.harvard.edu/abs/2021ApJ...916...96A}
}

@ARTICLE{Alazraki_1971_AA,
       author = {{Alazraki}, G. and {Couturier}, P.},
        title = "{Solar Wind Accejeration Caused by the Gradient of Alfven Wave Pressure}",
      journal = {\aap},
         year = 1971,
        month = aug,
       volume = {13},
        pages = {380},
       adsurl = {https://ui.adsabs.harvard.edu/abs/1971A&A....13..380A}
}

@ARTICLE{Altschuler_1969_SolPhys,
       author = {{Altschuler}, Martin D. and {Newkirk}, Gordon},
        title = "{Magnetic Fields and the Structure of the Solar Corona. I: Methods of Calculating Coronal Fields}",
      journal = {\solphys},
         year = 1969,
        month = sep,
       volume = {9},
       number = {1},
        pages = {131-149},
          doi = {10.1007/BF00145734},
       adsurl = {https://ui.adsabs.harvard.edu/abs/1969SoPh....9..131A}
}

@ARTICLE{Alvarado-Gomez_2016_AA,
       author = {{Alvarado-G{\'o}mez}, J.~D. and {Hussain}, G.~A.~J. and {Cohen}, O. and {Drake}, J.~J. and {Garraffo}, C. and {Grunhut}, J. and {Gombosi}, T.~I.},
        title = "{Simulating the environment around planet-hosting stars. I. Coronal structure}",
      journal = {\aap},
         year = 2016,
        month = apr,
       volume = {588},
          eid = {A28},
        pages = {A28},
          doi = {10.1051/0004-6361/201527832},
archivePrefix = {arXiv},
       eprint = {1601.04443},
 primaryClass = {astro-ph.SR},
       adsurl = {https://ui.adsabs.harvard.edu/abs/2016A&A...588A..28A}
}

@ARTICLE{Antiochos_2011_ApJ,
       author = {{Antiochos}, S.~K. and {Miki{\'c}}, Z. and {Titov}, V.~S. and {Lionello}, R. and {Linker}, J.~A.},
        title = "{A Model for the Sources of the Slow Solar Wind}",
      journal = {\apj},
         year = 2011,
        month = apr,
       volume = {731},
       number = {2},
          eid = {112},
        pages = {112},
          doi = {10.1088/0004-637X/731/2/112},
archivePrefix = {arXiv},
       eprint = {1102.3704},
 primaryClass = {astro-ph.SR},
       adsurl = {https://ui.adsabs.harvard.edu/abs/2011ApJ...731..112A}
}

@ARTICLE{Arge_2000_JGR,
       author = {{Arge}, C.~N. and {Pizzo}, V.~J.},
        title = "{Improvement in the prediction of solar wind conditions using near-real time solar magnetic field updates}",
      journal = {\jgr},
         year = 2000,
        month = may,
       volume = {105},
       number = {A5},
        pages = {10465-10480},
          doi = {10.1029/1999JA000262},
       adsurl = {https://ui.adsabs.harvard.edu/abs/2000JGR...10510465A}
}

@INPROCEEDINGS{Arge_2010_AIPC,
       author = {{Arge}, C. Nick and {Henney}, Carl J. and {Koller}, Josef and {Compeau}, C. Rich and {Young}, Shawn and {MacKenzie}, David and {Fay}, Alex and {Harvey}, John W.},
        title = "{Air Force Data Assimilative Photospheric Flux Transport (ADAPT) Model}",
    booktitle = {Twelfth International Solar Wind Conference},
         year = 2010,
       editor = {{Maksimovic}, M. and {Issautier}, K. and {Meyer-Vernet}, N. and {Moncuquet}, M. and {Pantellini}, F.},
       series = {American Institute of Physics Conference Series},
       volume = {1216},
        month = mar,
        pages = {343-346},
          doi = {10.1063/1.3395870},
       adsurl = {https://ui.adsabs.harvard.edu/abs/2010AIPC.1216..343A}
}

@INPROCEEDINGS{Arge_2013_AIPC,
       author = {{Arge}, C. Nick and {Henney}, Carl J. and {Hernandez}, Irene Gonzalez and {Toussaint}, W. Alex and {Koller}, Josef and {Godinez}, Humberto C.},
        title = "{Modeling the corona and solar wind using ADAPT maps that include far-side observations}",
    booktitle = {Solar Wind 13},
         year = 2013,
       editor = {{Zank}, Gary P. and {Borovsky}, Joe and {Bruno}, Roberto and {Cirtain}, Jonathan and {Cranmer}, Steve and {Elliott}, Heather and {Giacalone}, Joe and {Gonzalez}, Walter and {Li}, Gang and {Marsch}, Eckart and {Moebius}, Ebehard and {Pogorelov}, Nick and {Spann}, Jim and {Verkhoglyadova}, Olga},
       series = {American Institute of Physics Conference Series},
       volume = {1539},
        month = jun,
    publisher = {AIP},
        pages = {11-14},
          doi = {10.1063/1.4810977},
       adsurl = {https://ui.adsabs.harvard.edu/abs/2013AIPC.1539...11A}
}

@ARTICLE{Bale_2023_Nature,
       author = {{Bale}, S.~D. and {Drake}, J.~F. and {McManus}, M.~D. and {Desai}, M.~I. and {Badman}, S.~T. and {Larson}, D.~E. and {Swisdak}, M. and {Horbury}, T.~S. and {Raouafi}, N.~E. and {Phan}, T. and {Velli}, M. and {McComas}, D.~J. and {Cohen}, C.~M.~S. and {Mitchell}, D. and {Panasenco}, O. and {Kasper}, J.~C.},
        title = "{Interchange reconnection as the source of the fast solar wind within coronal holes}",
      journal = {\nat},
         year = 2023,
        month = jun,
       volume = {618},
       number = {7964},
        pages = {252-256},
          doi = {10.1038/s41586-023-05955-3},
archivePrefix = {arXiv},
       eprint = {2208.07932},
 primaryClass = {astro-ph.SR},
       adsurl = {https://ui.adsabs.harvard.edu/abs/2023Natur.618..252B}
}

@ARTICLE{Belcher_1971_ApJ,
       author = {{Belcher}, J.~W.},
        title = "{ALFV{\'E}NIC Wave Pressures and the Solar Wind}",
      journal = {\apj},
         year = 1971,
        month = sep,
       volume = {168},
        pages = {509},
          doi = {10.1086/151105},
       adsurl = {https://ui.adsabs.harvard.edu/abs/1971ApJ...168..509B}
}

@ARTICLE{Berger_1996_ApJ,
       author = {{Berger}, T.~E. and {Title}, A.~M.},
        title = "{On the Dynamics of Small-Scale Solar Magnetic Elements}",
      journal = {\apj},
         year = 1996,
        month = may,
       volume = {463},
        pages = {365},
          doi = {10.1086/177250},
       adsurl = {https://ui.adsabs.harvard.edu/abs/1996ApJ...463..365B}
}

@ARTICLE{Berger_2001_ApJ,
       author = {{Berger}, T.~E. and {Title}, A.~M.},
        title = "{On the Relation of G-Band Bright Points to the Photospheric Magnetic Field}",
      journal = {\apj},
         year = 2001,
        month = may,
       volume = {553},
       number = {1},
        pages = {449-469},
          doi = {10.1086/320663},
       adsurl = {https://ui.adsabs.harvard.edu/abs/2001ApJ...553..449B}
}

@ARTICLE{Boro_Saikia_2020_AandA,
       author = {{Boro Saikia}, S. and {Jin}, M. and {Johnstone}, C.~P. and {L{\"u}ftinger}, T. and {G{\"u}del}, M. and {Airapetian}, V.~S. and {Kislyakova}, K.~G. and {Folsom}, C.~P.},
        title = "{The solar wind from a stellar perspective. How do low-resolution data impact the determination of wind properties?}",
      journal = {\aap},
         year = 2020,
        month = mar,
       volume = {635},
          eid = {A178},
        pages = {A178},
          doi = {10.1051/0004-6361/201937107},
archivePrefix = {arXiv},
       eprint = {2002.10460},
 primaryClass = {astro-ph.SR},
       adsurl = {https://ui.adsabs.harvard.edu/abs/2020A&A...635A.178B}
}

@ARTICLE{Breu_2022_AA,
       author = {{Breu}, C. and {Peter}, H. and {Cameron}, R. and {Solanki}, S.~K. and {Przybylski}, D. and {Rempel}, M. and {Chitta}, L.~P.},
        title = "{A solar coronal loop in a box: Energy generation and heating}",
      journal = {\aap},
         year = 2022,
        month = feb,
       volume = {658},
          eid = {A45},
        pages = {A45},
          doi = {10.1051/0004-6361/202141451},
archivePrefix = {arXiv},
       eprint = {2112.11549},
 primaryClass = {astro-ph.SR},
       adsurl = {https://ui.adsabs.harvard.edu/abs/2022A&A...658A..45B}
}

@ARTICLE{Breu_2026_MNRAS,
       author = {{Breu}, C.~A. and {Pontin}, D.~I. and {Priest}, E. and {De Moortel}, I.},
        title = "{On the complex nature of coronal heating}",
      journal = {\mnras},
         year = 2026,
        month = jan,
       volume = {545},
       number = {3},
          eid = {staf2180},
        pages = {staf2180},
          doi = {10.1093/mnras/staf2180},
archivePrefix = {arXiv},
       eprint = {2512.17880},
 primaryClass = {astro-ph.SR},
       adsurl = {https://ui.adsabs.harvard.edu/abs/2026MNRAS.545f2180B}
}

@ARTICLE{Brott_2011_AandA,
       author = {{Brott}, I. and {de Mink}, S.~E. and {Cantiello}, M. and {Langer}, N. and {de Koter}, A. and {Evans}, C.~J. and {Hunter}, I. and {Trundle}, C. and {Vink}, J.~S.},
        title = "{Rotating massive main-sequence stars. I. Grids of evolutionary models and isochrones}",
      journal = {\aap},
         year = 2011,
        month = jun,
       volume = {530},
          eid = {A115},
        pages = {A115},
          doi = {10.1051/0004-6361/201016113},
archivePrefix = {arXiv},
       eprint = {1102.0530},
 primaryClass = {astro-ph.SR},
       adsurl = {https://ui.adsabs.harvard.edu/abs/2011A&A...530A.115B}
}

@ARTICLE{Bruls_1995_AandA,
       author = {{Bruls}, J.~H.~M.~J. and {Solanki}, S.~K.},
        title = "{Infrared lines as probes of solar magnetic features. IX. MgI 12{\ensuremath{\mu}}m diagnostics of solar plage.}",
      journal = {\aap},
         year = 1995,
        month = jan,
       volume = {293},
        pages = {240-251},
       adsurl = {https://ui.adsabs.harvard.edu/abs/1995A&A...293..240B}
}

@ARTICLE{Canet_2024_MNRAS,
       author = {{Canet}, Ada and {Varela}, Jacobo and {G{\'o}mez de Castro}, Ana I.},
        title = "{Stellar wind impact on early atmospheres around unmagnetized Earth-like planets}",
      journal = {\mnras},
         year = 2024,
        month = jun,
       volume = {531},
       number = {2},
        pages = {2626-2641},
          doi = {10.1093/mnras/stae1267},
archivePrefix = {arXiv},
       eprint = {2405.10641},
 primaryClass = {astro-ph.EP},
       adsurl = {https://ui.adsabs.harvard.edu/abs/2024MNRAS.531.2626C}
}

@ARTICLE{Carbone_2009_PhRvL,
       author = {{Carbone}, V. and {Marino}, R. and {Sorriso-Valvo}, L. and {Noullez}, A. and {Bruno}, R.},
        title = "{Scaling Laws of Turbulence and Heating of Fast Solar Wind: The Role of Density Fluctuations}",
      journal = {\prl},
         year = 2009,
        month = aug,
       volume = {103},
       number = {6},
          eid = {061102},
        pages = {061102},
          doi = {10.1103/PhysRevLett.103.061102},
archivePrefix = {arXiv},
       eprint = {1003.0533},
 primaryClass = {physics.space-ph},
       adsurl = {https://ui.adsabs.harvard.edu/abs/2009PhRvL.103f1102C}
}

@ARTICLE{Chandran_2019_JPP,
       author = {{Chandran}, Benjamin D.~G. and {Perez}, Jean C.},
        title = "{Reflection-driven magnetohydrodynamic turbulence in the solar atmosphere and solar wind}",
      journal = {Journal of Plasma Physics},
         year = 2019,
        month = aug,
       volume = {85},
       number = {4},
          eid = {905850409},
        pages = {905850409},
          doi = {10.1017/S0022377819000540},
archivePrefix = {arXiv},
       eprint = {1908.00880},
 primaryClass = {physics.space-ph},
       adsurl = {https://ui.adsabs.harvard.edu/abs/2019JPlPh..85d9009C}
}

@ARTICLE{Cho_2003_MNRAS,
       author = {{Cho}, Jungyeon and {Lazarian}, A.},
        title = "{Compressible magnetohydrodynamic turbulence: mode coupling, scaling relations, anisotropy, viscosity-damped regime and astrophysical implications}",
      journal = {\mnras},
     keywords = {MHD, turbulence, ISM: general, Astrophysics, Physics - Fluid Dynamics, Physics - Plasma Physics},
         year = 2003,
        month = oct,
       volume = {345},
       number = {12},
        pages = {325-339},
          doi = {10.1046/j.1365-8711.2003.06941.x},
archivePrefix = {arXiv},
       eprint = {astro-ph/0301062},
 primaryClass = {astro-ph},
       adsurl = {https://ui.adsabs.harvard.edu/abs/2003MNRAS.345..325C},
      _adsnote = {Provided by the SAO/NASA Astrophysics Data System}
}

@ARTICLE{Cohen_2011_MNRAS,
       author = {{Cohen}, O.},
        title = "{The independency of stellar mass-loss rates on stellar X-ray luminosity and activity level based on solar X-ray flux and solar wind observations}",
      journal = {\mnras},
         year = 2011,
        month = nov,
       volume = {417},
       number = {4},
        pages = {2592-2600},
          doi = {10.1111/j.1365-2966.2011.19428.x},
archivePrefix = {arXiv},
       eprint = {1107.2275},
 primaryClass = {astro-ph.SR},
       adsurl = {https://ui.adsabs.harvard.edu/abs/2011MNRAS.417.2592C}
}

@ARTICLE{Cranmer_2005_ApJ,
       author = {{Cranmer}, S.~R. and {van Ballegooijen}, A.~A.},
        title = "{On the Generation, Propagation, and Reflection of Alfv{\'e}n Waves from the Solar Photosphere to the Distant Heliosphere}",
      journal = {\apjs},
         year = 2005,
        month = feb,
       volume = {156},
       number = {2},
        pages = {265-293},
          doi = {10.1086/426507},
archivePrefix = {arXiv},
       eprint = {astro-ph/0410639},
 primaryClass = {astro-ph},
       adsurl = {https://ui.adsabs.harvard.edu/abs/2005ApJS..156..265C}
}

@ARTICLE{Cranmer_2007_ApJ,
       author = {{Cranmer}, Steven R. and {van Ballegooijen}, Adriaan A. and {Edgar}, Richard J.},
        title = "{Self-consistent Coronal Heating and Solar Wind Acceleration from Anisotropic Magnetohydrodynamic Turbulence}",
      journal = {\apjs},
         year = 2007,
        month = aug,
       volume = {171},
       number = {2},
        pages = {520-551},
          doi = {10.1086/518001},
archivePrefix = {arXiv},
       eprint = {astro-ph/0703333},
 primaryClass = {astro-ph},
       adsurl = {https://ui.adsabs.harvard.edu/abs/2007ApJS..171..520C}
}

@ARTICLE{Cranmer_2011_ApJ,
       author = {{Cranmer}, Steven R. and {Saar}, Steven H.},
        title = "{Testing a Predictive Theoretical Model for the Mass Loss Rates of Cool Stars}",
      journal = {\apj},
         year = 2011,
        month = nov,
       volume = {741},
       number = {1},
          eid = {54},
        pages = {54},
          doi = {10.1088/0004-637X/741/1/54},
archivePrefix = {arXiv},
       eprint = {1108.4369},
 primaryClass = {astro-ph.SR},
       adsurl = {https://ui.adsabs.harvard.edu/abs/2011ApJ...741...54C}
}

@ARTICLE{Dakeyo_2024_AandA,
       author = {{Dakeyo}, J. -B. and {Rouillard}, A.~P. and {R{\'e}ville}, V. and {D{\'e}moulin}, P. and {Maksimovic}, M. and {Chapiron}, A. and {Pinto}, R.~F. and {Louarn}, P.},
        title = "{Testing the flux tube expansion factor - solar wind speed relation with Solar Orbiter data}",
      journal = {\aap},
         year = 2024,
        month = nov,
       volume = {691},
          eid = {A77},
        pages = {A77},
          doi = {10.1051/0004-6361/202451272},
archivePrefix = {arXiv},
       eprint = {2408.06155},
 primaryClass = {astro-ph.SR},
       adsurl = {https://ui.adsabs.harvard.edu/abs/2024A&A...691A..77D}
}

@ARTICLE{Del_Zanna_2021_ApJ,
       author = {{Del Zanna}, G. and {Dere}, K.~P. and {Young}, P.~R. and {Landi}, E.},
        title = "{CHIANTI{\textemdash}An Atomic Database for Emission Lines. XVI. Version 10, Further Extensions}",
      journal = {\apj},
         year = 2021,
        month = mar,
       volume = {909},
       number = {1},
          eid = {38},
        pages = {38},
          doi = {10.3847/1538-4357/abd8ce},
archivePrefix = {arXiv},
       eprint = {2011.05211},
 primaryClass = {physics.atom-ph},
       adsurl = {https://ui.adsabs.harvard.edu/abs/2021ApJ...909...38D}
}

@ARTICLE{Dere_1997_AA,
       author = {{Dere}, K.~P. and {Landi}, E. and {Mason}, H.~E. and {Monsignori Fossi}, B.~C. and {Young}, P.~R.},
        title = "{CHIANTI - an atomic database for emission lines}",
      journal = {\aaps},
         year = 1997,
        month = oct,
       volume = {125},
        pages = {149-173},
          doi = {10.1051/aas:1997368},
       adsurl = {https://ui.adsabs.harvard.edu/abs/1997A&AS..125..149D}
}

@ARTICLE{Dmitruk_2002_ApJ,
       author = {{Dmitruk}, P. and {Matthaeus}, W.~H. and {Milano}, L.~J. and {Oughton}, S. and {Zank}, G.~P. and {Mullan}, D.~J.},
        title = "{Coronal Heating Distribution Due to Low-Frequency, Wave-driven Turbulence}",
      journal = {\apj},
         year = 2002,
        month = aug,
       volume = {575},
       number = {1},
        pages = {571-577},
          doi = {10.1086/341188},
archivePrefix = {arXiv},
       eprint = {astro-ph/0204347},
 primaryClass = {astro-ph},
       adsurl = {https://ui.adsabs.harvard.edu/abs/2002ApJ...575..571D}
}

@ARTICLE{Dong_2017_ApJ,
       author = {{Dong}, Chuanfei and {Lingam}, Manasvi and {Ma}, Yingjuan and {Cohen}, Ofer},
        title = "{Is Proxima Centauri b Habitable? A Study of Atmospheric Loss}",
      journal = {\apjl},
         year = 2017,
        month = mar,
       volume = {837},
       number = {2},
          eid = {L26},
        pages = {L26},
          doi = {10.3847/2041-8213/aa6438},
archivePrefix = {arXiv},
       eprint = {1702.04089},
 primaryClass = {astro-ph.EP},
       adsurl = {https://ui.adsabs.harvard.edu/abs/2017ApJ...837L..26D}
}

@ARTICLE{Dong_2022_SciA,
       author = {{Dong}, Chuanfei and {Wang}, Liang and {Huang}, Yi-Min and {Comisso}, Luca and {Sandstrom}, Timothy A. and {Bhattacharjee}, Amitava},
        title = "{Reconnection-driven energy cascade in magnetohydrodynamic turbulence}",
      journal = {Science Advances},
         year = 2022,
        month = dec,
       volume = {8},
       number = {49},
          eid = {eabn7627},
        pages = {eabn7627},
          doi = {10.1126/sciadv.abn7627},
archivePrefix = {arXiv},
       eprint = {2210.10736},
 primaryClass = {astro-ph.SR},
       adsurl = {https://ui.adsabs.harvard.edu/abs/2022SciA....8N7627D}
}

@ARTICLE{Downs_2016_ApJ,
       author = {{Downs}, Cooper and {Lionello}, Roberto and {Miki{\'c}}, Zoran and {Linker}, Jon A. and {Velli}, Marco},
        title = "{Closed-field Coronal Heating Driven by Wave Turbulence}",
      journal = {\apj},
         year = 2016,
        month = dec,
       volume = {832},
       number = {2},
          eid = {180},
        pages = {180},
          doi = {10.3847/0004-637X/832/2/180},
archivePrefix = {arXiv},
       eprint = {1610.02113},
 primaryClass = {astro-ph.SR},
       adsurl = {https://ui.adsabs.harvard.edu/abs/2016ApJ...832..180D}
}

@ARTICLE{Edwards_2006_ApJ,
       author = {{Edwards}, Suzan and {Fischer}, William and {Hillenbrand}, Lynne and {Kwan}, John},
        title = "{Probing T Tauri Accretion and Outflow with 1 Micron Spectroscopy}",
      journal = {\apj},
         year = 2006,
        month = jul,
       volume = {646},
       number = {1},
        pages = {319-341},
          doi = {10.1086/504832},
archivePrefix = {arXiv},
       eprint = {astro-ph/0604006},
 primaryClass = {astro-ph},
       adsurl = {https://ui.adsabs.harvard.edu/abs/2006ApJ...646..319E}
}

@ARTICLE{Erkal_2022_AandA,
       author = {{Erkal}, J. and {Manara}, C.~F. and {Schneider}, P.~C. and {Vincenzi}, M. and {Nisini}, B. and {Coffey}, D. and {Alcal{\'a}}, J.~M. and {Fedele}, D. and {Antoniucci}, S.},
        title = "{The He I {\ensuremath{\lambda}}10830 {\r{A}} line as a probe of winds and accretion in young stars in Lupus and Upper Scorpius}",
      journal = {\aap},
         year = 2022,
        month = oct,
       volume = {666},
          eid = {A188},
        pages = {A188},
          doi = {10.1051/0004-6361/202244254},
archivePrefix = {arXiv},
       eprint = {2208.02940},
 primaryClass = {astro-ph.SR},
       adsurl = {https://ui.adsabs.harvard.edu/abs/2022A&A...666A.188E}
}

@ARTICLE{Evensberget_2021_MNRAS,
       author = {{Evensberget}, D. and {Carter}, B.~D. and {Marsden}, S.~C. and {Brookshaw}, L. and {Folsom}, C.~P.},
        title = "{The winds of young Solar-type stars in the Hyades}",
      journal = {\mnras},
         year = 2021,
        month = sep,
       volume = {506},
       number = {2},
        pages = {2309-2335},
          doi = {10.1093/mnras/stab1696},
archivePrefix = {arXiv},
       eprint = {2106.04937},
 primaryClass = {astro-ph.SR},
       adsurl = {https://ui.adsabs.harvard.edu/abs/2021MNRAS.506.2309E}
}

@ARTICLE{Evensberget_2022_MNRAS,
       author = {{Evensberget}, D. and {Carter}, B.~D. and {Marsden}, S.~C. and {Brookshaw}, L. and {Folsom}, C.~P. and {Salmeron}, R.},
        title = "{The winds of young Solar-type stars in Coma Berenices and Hercules-Lyra}",
      journal = {\mnras},
         year = 2022,
        month = mar,
       volume = {510},
       number = {4},
        pages = {5226-5245},
          doi = {10.1093/mnras/stab3557},
archivePrefix = {arXiv},
       eprint = {2112.01445},
 primaryClass = {astro-ph.SR},
       adsurl = {https://ui.adsabs.harvard.edu/abs/2022MNRAS.510.5226E}
}

@ARTICLE{Fedun_2011_AnGeo,
       author = {{Fedun}, V. and {Shelyag}, S. and {Verth}, G. and {Mathioudakis}, M. and {Erd{\'e}lyi}, R.},
        title = "{MHD waves generated by high-frequency photospheric vortex motions}",
      journal = {Annales Geophysicae},
         year = 2011,
        month = jun,
       volume = {29},
       number = {6},
        pages = {1029-1035},
          doi = {10.5194/angeo-29-1029-2011},
       adsurl = {https://ui.adsabs.harvard.edu/abs/2011AnGeo..29.1029F}
}

@ARTICLE{Finley_2022_AA,
       author = {{Finley}, A.~J. and {Brun}, A.~S. and {Carlsson}, M. and {Szydlarski}, M. and {Hansteen}, V. and {Shoda}, M.},
        title = "{Stirring the base of the solar wind: On heat transfer and vortex formation}",
      journal = {\aap},
         year = 2022,
        month = sep,
       volume = {665},
          eid = {A118},
        pages = {A118},
          doi = {10.1051/0004-6361/202243947},
archivePrefix = {arXiv},
       eprint = {2207.02878},
 primaryClass = {astro-ph.SR},
       adsurl = {https://ui.adsabs.harvard.edu/abs/2022A&A...665A.118F}
}

@ARTICLE{Fisk_2003_JGR,
       author = {{Fisk}, L.~A.},
        title = "{Acceleration of the solar wind as a result of the reconnection of open magnetic flux with coronal loops}",
      journal = {Journal of Geophysical Research (Space Physics)},
         year = 2003,
        month = apr,
       volume = {108},
       number = {A4},
          eid = {1157},
        pages = {1157},
          doi = {10.1029/2002JA009284},
       adsurl = {https://ui.adsabs.harvard.edu/abs/2003JGRA..108.1157F}
}

@ARTICLE{Fujiki_2015_SoPh,
       author = {{Fujiki}, Ken'ichi and {Tokumaru}, Munetoshi and {Iju}, Tomoya and {Hakamada}, Kazuyuki and {Kojima}, Masayoshi},
        title = "{Relationship Between Solar-Wind Speed and Coronal Magnetic-Field Properties}",
      journal = {\solphys},
         year = 2015,
        month = sep,
       volume = {290},
       number = {9},
        pages = {2491-2505},
          doi = {10.1007/s11207-015-0742-8},
archivePrefix = {arXiv},
       eprint = {1507.03301},
 primaryClass = {astro-ph.SR},
       adsurl = {https://ui.adsabs.harvard.edu/abs/2015SoPh..290.2491F}
}

@ARTICLE{Fujimura_2009_ApJ,
       author = {{Fujimura}, D. and {Tsuneta}, S.},
        title = "{Properties of Magnetohydrodynamic Waves in the Solar Photosphere Obtained with Hinode}",
      journal = {\apj},
         year = 2009,
        month = sep,
       volume = {702},
       number = {2},
        pages = {1443-1457},
          doi = {10.1088/0004-637X/702/2/1443},
archivePrefix = {arXiv},
       eprint = {0907.3025},
 primaryClass = {astro-ph.SR},
       adsurl = {https://ui.adsabs.harvard.edu/abs/2009ApJ...702.1443F}
}

@ARTICLE{Gallet_2013_AA,
       author = {{Gallet}, F. and {Bouvier}, J.},
        title = "{Improved angular momentum evolution model for solar-like stars}",
      journal = {\aap},
         year = 2013,
        month = aug,
       volume = {556},
          eid = {A36},
        pages = {A36},
          doi = {10.1051/0004-6361/201321302},
archivePrefix = {arXiv},
       eprint = {1306.2130},
 primaryClass = {astro-ph.SR},
       adsurl = {https://ui.adsabs.harvard.edu/abs/2013A&A...556A..36G}
}

@ARTICLE{Gallet_2015_AA,
       author = {{Gallet}, F. and {Bouvier}, J.},
        title = "{Improved angular momentum evolution model for solar-like stars. II. Exploring the mass dependence}",
      journal = {\aap},
         year = 2015,
        month = may,
       volume = {577},
          eid = {A98},
        pages = {A98},
          doi = {10.1051/0004-6361/201525660},
archivePrefix = {arXiv},
       eprint = {1502.05801},
 primaryClass = {astro-ph.SR},
       adsurl = {https://ui.adsabs.harvard.edu/abs/2015A&A...577A..98G}
}

@ARTICLE{Garraffo_2016_ApJ,
       author = {{Garraffo}, C. and {Drake}, J.~J. and {Cohen}, O.},
        title = "{The Space Weather of Proxima Centauri b}",
      journal = {\apjl},
         year = 2016,
        month = dec,
       volume = {833},
       number = {1},
          eid = {L4},
        pages = {L4},
          doi = {10.3847/2041-8205/833/1/L4},
archivePrefix = {arXiv},
       eprint = {1609.09076},
 primaryClass = {astro-ph.SR},
       adsurl = {https://ui.adsabs.harvard.edu/abs/2016ApJ...833L...4G}
}

@ARTICLE{Goldreich_1995_ApJ,
       author = {{Goldreich}, P. and {Sridhar}, S.},
        title = "{Toward a Theory of Interstellar Turbulence. II. Strong Alfvenic Turbulence}",
      journal = {\apj},
         year = 1995,
        month = jan,
       volume = {438},
        pages = {763},
          doi = {10.1086/175121},
       adsurl = {https://ui.adsabs.harvard.edu/abs/1995ApJ...438..763G}
}

@ARTICLE{Goodman_2012_ApJ,
       author = {{Goodman}, Michael L. and {Judge}, Philip G.},
        title = "{Radiating Current Sheets in the Solar Chromosphere}",
      journal = {\apj},
         year = 2012,
        month = may,
       volume = {751},
       number = {1},
          eid = {75},
        pages = {75},
          doi = {10.1088/0004-637X/751/1/75},
archivePrefix = {arXiv},
       eprint = {1406.1211},
 primaryClass = {astro-ph.SR},
       adsurl = {https://ui.adsabs.harvard.edu/abs/2012ApJ...751...75G}
}

@ARTICLE{Gottlieb_2001_SIAMR,
       author = {{Gottlieb}, Sigal and {Shu}, Chi-Wang and {Tadmor}, Eitan},
        title = "{Strong Stability-Preserving High-Order Time Discretization Methods}",
      journal = {SIAM Review},
         year = 2001,
        month = jan,
       volume = {43},
       number = {1},
        pages = {89-112},
          doi = {10.1137/S003614450036757X},
       adsurl = {https://ui.adsabs.harvard.edu/abs/2001SIAMR..43...89G}
}

@ARTICLE{Gu_1997_ApJ,
       author = {{Gu}, Yeming and {Jefferies}, John T. and {Lindsey}, Charles and {Avrett}, E.~H.},
        title = "{A Stochastic Model of the Solar Atmosphere}",
      journal = {\apj},
         year = 1997,
        month = jul,
       volume = {484},
       number = {2},
        pages = {960-978},
          doi = {10.1086/304353},
       adsurl = {https://ui.adsabs.harvard.edu/abs/1997ApJ...484..960G}
}

@ARTICLE{Hansteen_1995_JGR,
       author = {{Hansteen}, Viggo H. and {Leer}, Egil},
        title = "{Coronal heating, densities, and temperatures and solar wind acceleration}",
      journal = {\jgr},
         year = 1995,
        month = nov,
       volume = {100},
       number = {A11},
        pages = {21577-21594},
          doi = {10.1029/95JA02300},
       adsurl = {https://ui.adsabs.harvard.edu/abs/1995JGR...10021577H}
}

@ARTICLE{Hickmann_2015_SoPh,
       author = {{Hickmann}, Kyle S. and {Godinez}, Humberto C. and {Henney}, Carl J. and {Arge}, C. Nick},
        title = "{Data Assimilation in the ADAPT Photospheric Flux Transport Model}",
      journal = {\solphys},
         year = 2015,
        month = apr,
       volume = {290},
       number = {4},
        pages = {1105-1118},
          doi = {10.1007/s11207-015-0666-3},
archivePrefix = {arXiv},
       eprint = {1410.6185},
 primaryClass = {math-ph},
       adsurl = {https://ui.adsabs.harvard.edu/abs/2015SoPh..290.1105H}
}

@ARTICLE{Hofner_2018_AARev,
       author = {{H{\"o}fner}, Susanne and {Olofsson}, Hans},
        title = "{Mass loss of stars on the asymptotic giant branch. Mechanisms, models and measurements}",
      journal = {\aapr},
         year = 2018,
        month = jan,
       volume = {26},
       number = {1},
          eid = {1},
        pages = {1},
          doi = {10.1007/s00159-017-0106-5},
       adsurl = {https://ui.adsabs.harvard.edu/abs/2018A&ARv..26....1H}
}

@ARTICLE{Hollweg_1982_SolPhys,
       author = {{Hollweg}, J.~V. and {Jackson}, S. and {Galloway}, D.},
        title = "{Alfven Waves in the Solar Atmospheres - Part Three - Nonlinear Waves on Open Flux Tubes}",
      journal = {\solphys},
         year = 1982,
        month = jan,
       volume = {75},
       number = {1-2},
        pages = {35-61},
          doi = {10.1007/BF00153458},
       adsurl = {https://ui.adsabs.harvard.edu/abs/1982SoPh...75...35H}
}

@ARTICLE{Hollweg_1986_JGR,
       author = {{Hollweg}, J.~V.},
        title = "{Transition region, corona, and solar wind in coronal holes}",
      journal = {\jgr},
         year = 1986,
        month = apr,
       volume = {91},
       number = {A4},
        pages = {4111-4125},
          doi = {10.1029/JA091iA04p04111},
       adsurl = {https://ui.adsabs.harvard.edu/abs/1986JGR....91.4111H}
}

@ARTICLE{Howes_2016_ApJ,
       author = {{Howes}, Gregory G.},
        title = "{The Dynamical Generation of Current Sheets in Astrophysical Plasma Turbulence}",
      journal = {\apjl},
         year = 2016,
        month = aug,
       volume = {827},
       number = {2},
          eid = {L28},
        pages = {L28},
          doi = {10.3847/2041-8205/827/2/L28},
archivePrefix = {arXiv},
       eprint = {1607.07465},
 primaryClass = {astro-ph.SR},
       adsurl = {https://ui.adsabs.harvard.edu/abs/2016ApJ...827L..28H}
}

@PHDTHESIS{Iijima_2016_PhD,
       author = {{Iijima}, Haruhisa},
        title = "{Numerical studies of solar chromospheric jets}",
       school = {Department of Earth and Planetary Science, School of Science, The
        University of Tokyo, Japan},
         year = 2016,
        month = mar,
       adsurl = {https://ui.adsabs.harvard.edu/abs/2016PhDT.........5I}
}

@ARTICLE{Iijima_2021_ApJ,
       author = {{Iijima}, Haruhisa and {Imada}, Shinsuke},
        title = "{A New Broadening Technique of the Numerically Unresolved Solar Transition Region and Its Effect on the Spectroscopic Synthesis Using Coronal Approximation}",
      journal = {\apj},
         year = 2021,
        month = aug,
       volume = {917},
       number = {2},
          eid = {65},
        pages = {65},
          doi = {10.3847/1538-4357/ac07a5},
archivePrefix = {arXiv},
       eprint = {2106.00864},
 primaryClass = {astro-ph.SR},
       adsurl = {https://ui.adsabs.harvard.edu/abs/2021ApJ...917...65I}
}

@ARTICLE{Iijima_2023_ApJ,
       author = {{Iijima}, Haruhisa and {Matsumoto}, Takuma and {Hotta}, Hideyuki and {Imada}, Shinsuke},
        title = "{A Comprehensive Simulation of Solar Wind Formation from the Solar Interior: Significant Cross-field Energy Transport by Interchange Reconnection near the Sun}",
      journal = {\apjl},
         year = 2023,
        month = jul,
       volume = {951},
       number = {2},
          eid = {L47},
        pages = {L47},
          doi = {10.3847/2041-8213/acdde0},
archivePrefix = {arXiv},
       eprint = {2306.17324},
 primaryClass = {astro-ph.SR},
       adsurl = {https://ui.adsabs.harvard.edu/abs/2023ApJ...951L..47I}
}

@ARTICLE{Ishikawa_2021_ScienceAdvances,
       author = {{Ishikawa}, Ryohko and {Bueno}, Javier Trujillo and {del Pino Alem{\'a}n}, Tanaus{\'u} and {Okamoto}, Takenori J. and {McKenzie}, David E. and {Auch{\`e}re}, Fr{\'e}d{\'e}ric and {Kano}, Ryouhei and {Song}, Donguk and {Yoshida}, Masaki and {Rachmeler}, Laurel A. and {Kobayashi}, Ken and {Hara}, Hirohisa and {Kubo}, Masahito and {Narukage}, Noriyuki and {Sakao}, Taro and {Shimizu}, Toshifumi and {Suematsu}, Yoshinori and {Bethge}, Christian and {De Pontieu}, Bart and {Dalda}, Alberto Sainz and {Vigil}, Genevieve D. and {Winebarger}, Amy and {Ballester}, Ernest Alsina and {Belluzzi}, Luca and {{\v{S}}t{\v{e}}p{\'a}n}, Ji{\v{r}}{\'\i} and {Ramos}, Andr{\'e}s Asensio and {Carlsson}, Mats and {Leenaarts}, Jorrit},
        title = "{Mapping solar magnetic fields from the photosphere to the base of the corona}",
      journal = {Science Advances},
         year = 2021,
        month = feb,
       volume = {7},
       number = {8},
        pages = {eabe8406},
          doi = {10.1126/sciadv.abe8406},
archivePrefix = {arXiv},
       eprint = {2103.01583},
 primaryClass = {astro-ph.SR},
       adsurl = {https://ui.adsabs.harvard.edu/abs/2021SciA....7.8406I}
}

@ARTICLE{Jahandar_2025_ApJ,
       author = {{Jahandar}, Farbod and {Doyon}, Ren{\'e} and {Artigau}, {\'E}tienne and {Cook}, Neil J. and {Cadieux}, Charles and {Donati}, Jean-Fran{\c{c}}ois and {Cowan}, Nicolas B. and {Cloutier}, Ryan and {Pelletier}, Stefan and {Alves-Brito}, Alan and {Martins}, Jorge H.~C. and {Shang}, Hsien and {Carmona}, Andr{\'e}s},
        title = "{Chemical Fingerprints of M Dwarfs: High-resolution Spectroscopy on 31 M Dwarfs with SPIRou}",
      journal = {\apj},
         year = 2025,
        month = jan,
       volume = {978},
       number = {2},
          eid = {154},
        pages = {154},
          doi = {10.3847/1538-4357/ad91a0},
archivePrefix = {arXiv},
       eprint = {2411.07377},
 primaryClass = {astro-ph.SR},
       adsurl = {https://ui.adsabs.harvard.edu/abs/2025ApJ...978..154J}
}

@ARTICLE{Jardine_2019_MNRAS,
       author = {{Jardine}, Moira and {Collier Cameron}, Andrew},
        title = "{Slingshot prominences: nature's wind gauges}",
      journal = {\mnras},
         year = 2019,
        month = jan,
       volume = {482},
       number = {3},
        pages = {2853-2860},
          doi = {10.1093/mnras/sty2872},
archivePrefix = {arXiv},
       eprint = {1810.09319},
 primaryClass = {astro-ph.SR},
       adsurl = {https://ui.adsabs.harvard.edu/abs/2019MNRAS.482.2853J}
}

@ARTICLE{Kannan_2024_MNRAS,
       author = {{Kannan}, Arjun and {Yadav}, Nitin},
        title = "{Vortex dynamics in various solar magnetic field configurations}",
      journal = {\mnras},
         year = 2024,
        month = sep,
       volume = {533},
       number = {3},
        pages = {3611-3622},
          doi = {10.1093/mnras/stae1990},
archivePrefix = {arXiv},
       eprint = {2408.08225},
 primaryClass = {astro-ph.SR},
       adsurl = {https://ui.adsabs.harvard.edu/abs/2024MNRAS.533.3611K}
}

@ARTICLE{Kavanagh_2021_MNRAS,
       author = {{Kavanagh}, Robert D. and {Vidotto}, Aline A. and {Klein}, Baptiste and {Jardine}, Moira M. and {Donati}, Jean-Fran{\c{c}}ois and {{\'O} Fionnag{\'a}in}, D{\'u}alta},
        title = "{Planet-induced radio emission from the coronae of M dwarfs: the case of Prox Cen and AU Mic}",
      journal = {\mnras},
         year = 2021,
        month = jun,
       volume = {504},
       number = {1},
        pages = {1511-1518},
          doi = {10.1093/mnras/stab929},
archivePrefix = {arXiv},
       eprint = {2103.16318},
 primaryClass = {astro-ph.SR},
       adsurl = {https://ui.adsabs.harvard.edu/abs/2021MNRAS.504.1511K}
}

@ARTICLE{Khomenko_2012_ApJ,
       author = {{Khomenko}, E. and {Collados}, M.},
        title = "{Heating of the Magnetized Solar Chromosphere by Partial Ionization Effects}",
      journal = {\apj},
         year = 2012,
        month = mar,
       volume = {747},
       number = {2},
          eid = {87},
        pages = {87},
          doi = {10.1088/0004-637X/747/2/87},
archivePrefix = {arXiv},
       eprint = {1112.3374},
 primaryClass = {astro-ph.SR},
       adsurl = {https://ui.adsabs.harvard.edu/abs/2012ApJ...747...87K}
}

@ARTICLE{Khomenko_2014_PhPl,
       author = {{Khomenko}, E. and {Collados}, M. and {D{\'\i}az}, A. and {Vitas}, N.},
        title = "{Fluid description of multi-component solar partially ionized plasma}",
      journal = {Physics of Plasmas},
         year = 2014,
        month = sep,
       volume = {21},
       number = {9},
          eid = {092901},
        pages = {092901},
          doi = {10.1063/1.4894106},
archivePrefix = {arXiv},
       eprint = {1408.1871},
 primaryClass = {astro-ph.SR},
       adsurl = {https://ui.adsabs.harvard.edu/abs/2014PhPl...21i2901K}
}

@ARTICLE{Kislyakova_2024_NatAs,
       author = {{Kislyakova}, K.~G. and {G{\"u}del}, M. and {Koutroumpa}, D. and {Carter}, J.~A. and {Lisse}, C.~M. and {Boro Saikia}, S.},
        title = "{X-ray detection of astrospheres around three main-sequence stars and their mass-loss rates}",
      journal = {Nature Astronomy},
         year = 2024,
        month = may,
       volume = {8},
        pages = {596-605},
          doi = {10.1038/s41550-024-02222-x},
archivePrefix = {arXiv},
       eprint = {2404.14980},
 primaryClass = {astro-ph.SR},
       adsurl = {https://ui.adsabs.harvard.edu/abs/2024NatAs...8..596K}
}

@ARTICLE{Kochukhov_2020_AA,
       author = {{Kochukhov}, O. and {Hackman}, T. and {Lehtinen}, J.~J. and {Wehrhahn}, A.},
        title = "{Hidden magnetic fields of young suns}",
      journal = {\aap},
         year = 2020,
        month = mar,
       volume = {635},
          eid = {A142},
        pages = {A142},
          doi = {10.1051/0004-6361/201937185},
archivePrefix = {arXiv},
       eprint = {2002.10469},
 primaryClass = {astro-ph.SR},
       adsurl = {https://ui.adsabs.harvard.edu/abs/2020A&A...635A.142K}
}

@ARTICLE{Kopp_1976_SolPhys,
       author = {{Kopp}, R.~A. and {Holzer}, T.~E.},
        title = "{Dynamics of coronal hole regions. I. Steady polytropic flows with multiple critical points.}",
      journal = {\solphys},
         year = 1976,
        month = jul,
       volume = {49},
       number = {1},
        pages = {43-56},
          doi = {10.1007/BF00221484},
       adsurl = {https://ui.adsabs.harvard.edu/abs/1976SoPh...49...43K}
}

@ARTICLE{Kudoh_1999_ApJ,
       author = {{Kudoh}, Takahiro and {Shibata}, Kazunari},
        title = "{Alfv{\'e}n Wave Model of Spicules and Coronal Heating}",
      journal = {\apj},
         year = 1999,
        month = mar,
       volume = {514},
       number = {1},
        pages = {493-505},
          doi = {10.1086/306930},
       adsurl = {https://ui.adsabs.harvard.edu/abs/1999ApJ...514..493K}
}

@ARTICLE{Kudritzki_2000_ARAA,
       author = {{Kudritzki}, Rolf-Peter and {Puls}, Joachim},
        title = "{Winds from Hot Stars}",
      journal = {\araa},
         year = 2000,
        month = jan,
       volume = {38},
        pages = {613-666},
          doi = {10.1146/annurev.astro.38.1.613},
       adsurl = {https://ui.adsabs.harvard.edu/abs/2000ARA&A..38..613K}
}

@ARTICLE{Kuniyoshi_2023_ApJ,
       author = {{Kuniyoshi}, Hidetaka and {Shoda}, Munehito and {Iijima}, Haruhisa and {Yokoyama}, Takaaki},
        title = "{Magnetic Tornado Properties: A Substantial Contribution to the Solar Coronal Heating via Efficient Energy Transfer}",
      journal = {\apj},
         year = 2023,
        month = may,
       volume = {949},
       number = {1},
          eid = {8},
        pages = {8},
          doi = {10.3847/1538-4357/accbb8},
archivePrefix = {arXiv},
       eprint = {2304.03010},
 primaryClass = {astro-ph.SR},
       adsurl = {https://ui.adsabs.harvard.edu/abs/2023ApJ...949....8K}
}

@ARTICLE{Kuniyoshi_2024_ApJ,
       author = {{Kuniyoshi}, Hidetaka and {Bose}, Souvik and {Yokoyama}, Takaaki},
        title = "{Comprehensive Synthesis of Magnetic Tornado: Cospatial Incidence of Chromospheric Swirls and Extreme-ultraviolet Brightening}",
      journal = {\apjl},
         year = 2024,
        month = jul,
       volume = {969},
       number = {2},
          eid = {L34},
        pages = {L34},
          doi = {10.3847/2041-8213/ad5a0e},
archivePrefix = {arXiv},
       eprint = {2404.18892},
 primaryClass = {astro-ph.SR},
       adsurl = {https://ui.adsabs.harvard.edu/abs/2024ApJ...969L..34K}
}

@ARTICLE{Kuniyoshi_2025_ApJ,
       author = {{Kuniyoshi}, Hidetaka and {Imada}, Shinsuke and {Yokoyama}, Takaaki},
        title = "{A Unified Picture of Swirl-driven Solar Coronal Heating: Magnetic Energy Supply and Dissipation}",
      journal = {\apjl},
         year = 2025,
        month = sep,
       volume = {990},
       number = {2},
          eid = {L71},
        pages = {L71},
          doi = {10.3847/2041-8213/adfc61},
archivePrefix = {arXiv},
       eprint = {2508.14459},
 primaryClass = {astro-ph.SR},
       adsurl = {https://ui.adsabs.harvard.edu/abs/2025ApJ...990L..71K}
}

@ARTICLE{Kuwan_2007_ApJ,
       author = {{Kwan}, John and {Edwards}, Suzan and {Fischer}, William},
        title = "{Modeling T Tauri Winds from He I {\ensuremath{\lambda}}10830 Profiles}",
      journal = {\apj},
         year = 2007,
        month = mar,
       volume = {657},
       number = {2},
        pages = {897-915},
          doi = {10.1086/511057},
archivePrefix = {arXiv},
       eprint = {astro-ph/0611585},
 primaryClass = {astro-ph},
       adsurl = {https://ui.adsabs.harvard.edu/abs/2007ApJ...657..897K}
}

@BOOK{Lamers_1999_book,
       author = {{Lamers}, Henny J.~G.~L.~M. and {Cassinelli}, Joseph P.},
        title = "{Introduction to Stellar Winds}",
         year = 1999,
       adsurl = {https://ui.adsabs.harvard.edu/abs/1999isw..book.....L}
}

@ARTICLE{Leer_1980_JGR,
       author = {{Leer}, E. and {Holzer}, T.~E.},
        title = "{Energy addition in the solar wind.}",
      journal = {\jgr},
         year = 1980,
        month = sep,
       volume = {85},
        pages = {4681-4688},
          doi = {10.1029/JA085iA09p04681},
       adsurl = {https://ui.adsabs.harvard.edu/abs/1980JGR....85.4681L}
}

@ARTICLE{Lionello_2014_ApJ,
       author = {{Lionello}, Roberto and {Velli}, Marco and {Downs}, Cooper and {Linker}, Jon A. and {Miki{\'c}}, Zoran and {Verdini}, Andrea},
        title = "{Validating a Time-dependent Turbulence-driven Model of the Solar Wind}",
      journal = {\apj},
         year = 2014,
        month = apr,
       volume = {784},
       number = {2},
          eid = {120},
        pages = {120},
          doi = {10.1088/0004-637X/784/2/120},
archivePrefix = {arXiv},
       eprint = {1402.4188},
 primaryClass = {astro-ph.SR},
       adsurl = {https://ui.adsabs.harvard.edu/abs/2014ApJ...784..120L}
}

@ARTICLE{Magyar_2017_NatSR,
       author = {{Magyar}, Norbert and {Van Doorsselaere}, Tom and {Goossens}, Marcel},
        title = "{Generalized phase mixing: Turbulence-like behaviour from unidirectionally propagating MHD waves}",
      journal = {Scientific Reports},
         year = 2017,
        month = nov,
       volume = {7},
          eid = {14820},
        pages = {14820},
          doi = {10.1038/s41598-017-13660-1},
archivePrefix = {arXiv},
       eprint = {1702.02346},
 primaryClass = {astro-ph.SR},
       adsurl = {https://ui.adsabs.harvard.edu/abs/2017NatSR...714820M}
}

@ARTICLE{Martinez_Sykora_2017_Sci,
       author = {{Mart{\'\i}nez-Sykora}, J. and {De Pontieu}, B. and {Hansteen}, V.~H. and {Rouppe van der Voort}, L. and {Carlsson}, M. and {Pereira}, T.~M.~D.},
        title = "{On the generation of solar spicules and Alfv{\'e}nic waves}",
      journal = {Science},
         year = 2017,
        month = jun,
       volume = {356},
       number = {6344},
        pages = {1269-1272},
          doi = {10.1126/science.aah5412},
archivePrefix = {arXiv},
       eprint = {1710.07559},
 primaryClass = {astro-ph.SR},
       adsurl = {https://ui.adsabs.harvard.edu/abs/2017Sci...356.1269M}
}

@ARTICLE{Matsuoka_2024_ApJ,
       author = {{Matsuoka}, Masato and {Suzuki}, Takeru K. and {Tokuno}, Takato and {Kakiuchi}, Kensuke},
        title = "{Effect of Magnetic Diffusion in the Chromosphere on the Solar Wind}",
      journal = {\apj},
         year = 2024,
        month = jul,
       volume = {970},
       number = {1},
          eid = {16},
        pages = {16},
          doi = {10.3847/1538-4357/ad496b},
archivePrefix = {arXiv},
       eprint = {2403.18409},
 primaryClass = {astro-ph.SR},
       adsurl = {https://ui.adsabs.harvard.edu/abs/2024ApJ...970...16M}
}

@ARTICLE{Matsumoto_2019_PASJ,
       author = {{Matsumoto}, Yosuke and {Asahina}, Yuta and {Kudoh}, Yuki and {Kawashima}, Tomohisa and {Matsumoto}, Jin and {Takahashi}, Hiroyuki R. and {Minoshima}, Takashi and {Zenitani}, Seiji and {Miyoshi}, Takahiro and {Matsumoto}, Ryoji},
        title = "{Magnetohydrodynamic simulation code CANS+: Assessments and applications}",
      journal = {\pasj},
         year = 2019,
        month = aug,
       volume = {71},
       number = {4},
          eid = {83},
        pages = {83},
          doi = {10.1093/pasj/psz064},
archivePrefix = {arXiv},
       eprint = {1611.01775},
 primaryClass = {astro-ph.IM},
       adsurl = {https://ui.adsabs.harvard.edu/abs/2019PASJ...71...83M}
}

@ARTICLE{Matsumoto_2021_MNRAS,
       author = {{Matsumoto}, Takuma},
        title = "{Full compressible 3D MHD simulation of solar wind}",
      journal = {\mnras},
         year = 2021,
        month = jan,
       volume = {500},
       number = {4},
        pages = {4779-4787},
          doi = {10.1093/mnras/staa3533},
archivePrefix = {arXiv},
       eprint = {2009.03770},
 primaryClass = {astro-ph.SR},
       adsurl = {https://ui.adsabs.harvard.edu/abs/2021MNRAS.500.4779M}
}

@ARTICLE{McGregor_2011_JGRA,
       author = {{McGregor}, S.~L. and {Hughes}, W.~J. and {Arge}, C.~N. and {Odstrcil}, D. and {Schwadron}, N.~A.},
        title = "{The radial evolution of solar wind speeds}",
      journal = {Journal of Geophysical Research (Space Physics)},
         year = 2011,
        month = mar,
       volume = {116},
       number = {A3},
          eid = {A03106},
        pages = {A03106},
          doi = {10.1029/2010JA016006},
       adsurl = {https://ui.adsabs.harvard.edu/abs/2011JGRA..116.3106M}
}

@ARTICLE{Meyer_2012_MNRAS,
       author = {{Meyer}, Chad D. and {Balsara}, Dinshaw S. and {Aslam}, Tariq D.},
        title = "{A second-order accurate Super TimeStepping formulation for anisotropic thermal conduction}",
      journal = {\mnras},
         year = 2012,
        month = may,
       volume = {422},
       number = {3},
        pages = {2102-2115},
          doi = {10.1111/j.1365-2966.2012.20744.x},
       adsurl = {https://ui.adsabs.harvard.edu/abs/2012MNRAS.422.2102M}
}

@ARTICLE{Meyer_2014_JCP,
       author = {{Meyer}, Chad D. and {Balsara}, Dinshaw S. and {Aslam}, Tariq D.},
        title = "{A stabilized Runge-Kutta-Legendre method for explicit super-time-stepping of parabolic and mixed equations}",
      journal = {Journal of Computational Physics},
         year = 2014,
        month = jan,
       volume = {257},
        pages = {594-626},
          doi = {10.1016/j.jcp.2013.08.021},
       adsurl = {https://ui.adsabs.harvard.edu/abs/2014JCoPh.257..594M}
}

@ARTICLE{Mignone_2010_JCoPh,
       author = {{Mignone}, Andrea and {Tzeferacos}, Petros and {Bodo}, Gianluigi},
        title = "{High-order conservative finite difference GLM-MHD schemes for cell-centered MHD}",
      journal = {Journal of Computational Physics},
         year = 2010,
        month = aug,
       volume = {229},
       number = {17},
        pages = {5896-5920},
          doi = {10.1016/j.jcp.2010.04.013},
archivePrefix = {arXiv},
       eprint = {1001.2832},
 primaryClass = {astro-ph.HE},
       adsurl = {https://ui.adsabs.harvard.edu/abs/2010JCoPh.229.5896M}
}

@ARTICLE{Miyoshi_2005_JCP,
       author = {{Miyoshi}, Takahiro and {Kusano}, Kanya},
        title = "{A multi-state HLL approximate Riemann solver for ideal magnetohydrodynamics}",
      journal = {Journal of Computational Physics},
         year = 2005,
        month = sep,
       volume = {208},
       number = {1},
        pages = {315-344},
          doi = {10.1016/j.jcp.2005.02.017},
       adsurl = {https://ui.adsabs.harvard.edu/abs/2005JCoPh.208..315M}
}

@ARTICLE{Morton_2013_ApJ,
       author = {{Morton}, R.~J. and {Verth}, G. and {Fedun}, V. and {Shelyag}, S. and {Erd{\'e}lyi}, R.},
        title = "{Evidence for the Photospheric Excitation of Incompressible Chromospheric Waves}",
      journal = {\apj},
         year = 2013,
        month = may,
       volume = {768},
       number = {1},
          eid = {17},
        pages = {17},
          doi = {10.1088/0004-637X/768/1/17},
archivePrefix = {arXiv},
       eprint = {1303.2356},
 primaryClass = {astro-ph.SR},
       adsurl = {https://ui.adsabs.harvard.edu/abs/2013ApJ...768...17M}
}

@ARTICLE{Odstrcil_2003_AdSpR,
       author = {{Odstrcil}, D.},
        title = "{Modeling 3-D solar wind structure}",
      journal = {Advances in Space Research},
         year = 2003,
        month = aug,
       volume = {32},
       number = {4},
        pages = {497-506},
          doi = {10.1016/S0273-1177(03)00332-6},
       adsurl = {https://ui.adsabs.harvard.edu/abs/2003AdSpR..32..497O}
}

@ARTICLE{Passenger_2018_AandA,
       author = {{Passegger}, V.~M. and {Reiners}, A. and {Jeffers}, S.~V. and {Wende-von Berg}, S. and {Sch{\"o}fer}, P. and {Caballero}, J.~A. and {Schweitzer}, A. and {Amado}, P.~J. and {B{\'e}jar}, V.~J.~S. and {Cort{\'e}s-Contreras}, M. and {Hatzes}, A.~P. and {K{\"u}rster}, M. and {Montes}, D. and {Pedraz}, S. and {Quirrenbach}, A. and {Ribas}, I. and {Seifert}, W.},
        title = "{The CARMENES search for exoplanets around M dwarfs. Photospheric parameters of target stars from high-resolution spectroscopy}",
      journal = {\aap},
         year = 2018,
        month = jul,
       volume = {615},
          eid = {A6},
        pages = {A6},
          doi = {10.1051/0004-6361/201732312},
archivePrefix = {arXiv},
       eprint = {1802.02946},
 primaryClass = {astro-ph.SR},
       adsurl = {https://ui.adsabs.harvard.edu/abs/2018A&A...615A...6P}
}

@ARTICLE{Perri_2022_ApJ,
       author = {{Perri}, Barbara and {Leitner}, Peter and {Brchnelova}, Michaela and {Baratashvili}, Tinatin and {Ku{\'z}ma}, B{\l}a{\.z}ej and {Zhang}, Fan and {Lani}, Andrea and {Poedts}, Stefaan},
        title = "{COCONUT, a Novel Fast-converging MHD Model for Solar Corona Simulations: I. Benchmarking and Optimization of Polytropic Solutions}",
      journal = {\apj},
         year = 2022,
        month = sep,
       volume = {936},
       number = {1},
          eid = {19},
        pages = {19},
          doi = {10.3847/1538-4357/ac7237},
archivePrefix = {arXiv},
       eprint = {2205.03341},
 primaryClass = {astro-ph.SR},
       adsurl = {https://ui.adsabs.harvard.edu/abs/2022ApJ...936...19P}
}

@ARTICLE{Pognan_2018_ApJ,
       author = {{Pognan}, Quentin and {Garraffo}, Cecilia and {Cohen}, Ofer and {Drake}, Jeremy J.},
        title = "{The Solar Wind Environment in Time}",
      journal = {\apj},
         year = 2018,
        month = mar,
       volume = {856},
       number = {1},
          eid = {53},
        pages = {53},
          doi = {10.3847/1538-4357/aaaebb},
archivePrefix = {arXiv},
       eprint = {1802.05153},
 primaryClass = {astro-ph.SR},
       adsurl = {https://ui.adsabs.harvard.edu/abs/2018ApJ...856...53P}
}

@ARTICLE{Pomoell_2018_JSWSC,
       author = {{Pomoell}, Jens and {Poedts}, S.},
        title = "{EUHFORIA: European heliospheric forecasting information asset}",
      journal = {Journal of Space Weather and Space Climate},
         year = 2018,
        month = jun,
       volume = {8},
          eid = {A35},
        pages = {A35},
          doi = {10.1051/swsc/2018020},
       adsurl = {https://ui.adsabs.harvard.edu/abs/2018JSWSC...8A..35P}
}

@ARTICLE{Riley_2011_SoPh,
       author = {{Riley}, P. and {Lionello}, R.},
        title = "{Mapping Solar Wind Streams from the Sun to 1 AU: A Comparison of Techniques}",
      journal = {\solphys},
         year = 2011,
        month = jun,
       volume = {270},
       number = {2},
        pages = {575-592},
          doi = {10.1007/s11207-011-9766-x},
       adsurl = {https://ui.adsabs.harvard.edu/abs/2011SoPh..270..575R}
}

@ARTICLE{Roberts_2022_JGRA,
       author = {{Roberts}, O.~W. and {Alexandrova}, O. and {Sorriso-Valvo}, L. and {V{\"o}r{\"o}s}, Z. and {Nakamura}, R. and {Fischer}, D. and {Varsani}, A. and {Escoubet}, C. Philippe and {Volwerk}, M. and {Canu}, P. and {Lion}, S. and {Yearby}, K.},
        title = "{Scale-Dependent Kurtosis of Magnetic Field Fluctuations in the Solar Wind: A Multi-Scale Study With Cluster 2003-2015}",
      journal = {Journal of Geophysical Research (Space Physics)},
         year = 2022,
        month = sep,
       volume = {127},
       number = {9},
          eid = {e29483},
        pages = {e29483},
          doi = {10.1029/2021JA029483},
       adsurl = {https://ui.adsabs.harvard.edu/abs/2022JGRA..12729483R}
}

@ARTICLE{Rodriguez_Monoz_2019_AandA,
       author = {{Rodr{\'\i}guez-Mozos}, J.~M. and {Moya}, A.},
        title = "{Erosion of an exoplanetary atmosphere caused by stellar winds}",
      journal = {\aap},
         year = 2019,
        month = oct,
       volume = {630},
          eid = {A52},
        pages = {A52},
          doi = {10.1051/0004-6361/201935543},
archivePrefix = {arXiv},
       eprint = {1908.06695},
 primaryClass = {astro-ph.EP},
       adsurl = {https://ui.adsabs.harvard.edu/abs/2019A&A...630A..52R}
}

@ARTICLE{Sachdeva_2019_ApJ,
       author = {{Sachdeva}, Nishtha and {van der Holst}, Bart and {Manchester}, Ward B. and {T{\'o}th}, Gabor and {Chen}, Yuxi and {Lloveras}, Diego G. and {V{\'a}squez}, Alberto M. and {Lamy}, Philippe and {Wojak}, Julien and {Jackson}, Bernard V. and {Yu}, Hsiu-Shan and {Henney}, Carl J.},
        title = "{Validation of the Alfv{\'e}n Wave Solar Atmosphere Model (AWSoM) with Observations from the Low Corona to 1 au}",
      journal = {\apj},
         year = 2019,
        month = dec,
       volume = {887},
       number = {1},
          eid = {83},
        pages = {83},
          doi = {10.3847/1538-4357/ab4f5e},
archivePrefix = {arXiv},
       eprint = {1910.08110},
 primaryClass = {astro-ph.SR},
       adsurl = {https://ui.adsabs.harvard.edu/abs/2019ApJ...887...83S}
}

@ARTICLE{Sakaue_2021_ApJ,
       author = {{Sakaue}, Takahito and {Shibata}, Kazunari},
        title = "{An M Dwarf's Chromosphere, Corona, and Wind Connection via Nonlinear Alfv{\'e}n Waves}",
      journal = {\apj},
         year = 2021,
        month = sep,
       volume = {919},
       number = {1},
          eid = {29},
        pages = {29},
          doi = {10.3847/1538-4357/ac0e34},
archivePrefix = {arXiv},
       eprint = {2106.12752},
 primaryClass = {astro-ph.SR},
       adsurl = {https://ui.adsabs.harvard.edu/abs/2021ApJ...919...29S}
}

@ARTICLE{Schatten_1969_SolPhys,
       author = {{Schatten}, Kenneth H. and {Wilcox}, John M. and {Ness}, Norman F.},
        title = "{A model of interplanetary and coronal magnetic fields}",
      journal = {\solphys},
         year = 1969,
        month = mar,
       volume = {6},
       number = {3},
        pages = {442-455},
          doi = {10.1007/BF00146478},
       adsurl = {https://ui.adsabs.harvard.edu/abs/1969SoPh....6..442S}
}

@ARTICLE{Shebalin_1983_JPP,
       author = {{Shebalin}, J.~V. and {Matthaeus}, W.~H. and {Montgomery}, D.},
        title = "{Anisotropy in MHD turbulence due to a mean magnetic field}",
      journal = {Journal of Plasma Physics},
         year = 1983,
        month = jun,
       volume = {29},
       number = {3},
        pages = {525-547},
          doi = {10.1017/S0022377800000933},
       adsurl = {https://ui.adsabs.harvard.edu/abs/1983JPlPh..29..525S}
}

@ARTICLE{Shelyag_2012_ApJ,
       author = {{Shelyag}, S. and {Mathioudakis}, M. and {Keenan}, F.~P.},
        title = "{Mechanisms for MHD Poynting Flux Generation in Simulations of Solar Photospheric Magnetoconvection}",
      journal = {\apjl},
         year = 2012,
        month = jul,
       volume = {753},
       number = {1},
          eid = {L22},
        pages = {L22},
          doi = {10.1088/2041-8205/753/1/L22},
archivePrefix = {arXiv},
       eprint = {1206.0030},
 primaryClass = {astro-ph.SR},
       adsurl = {https://ui.adsabs.harvard.edu/abs/2012ApJ...753L..22S}
}

@ARTICLE{Shimizu_2022_ApJ,
       author = {{Shimizu}, Kimihiko and {Shoda}, Munehito and {Suzuki}, Takeru K.},
        title = "{Role of Longitudinal Waves in Alfv{\'e}n-wave-driven Solar Wind}",
      journal = {\apj},
         year = 2022,
        month = may,
       volume = {931},
       number = {1},
          eid = {37},
        pages = {37},
          doi = {10.3847/1538-4357/ac66d7},
archivePrefix = {arXiv},
       eprint = {2203.15280},
 primaryClass = {astro-ph.SR},
       adsurl = {https://ui.adsabs.harvard.edu/abs/2022ApJ...931...37S}
}

@ARTICLE{Shiota_2014_SpaceWeather,
       author = {{Shiota}, D. and {Kataoka}, R. and {Miyoshi}, Y. and {Hara}, T. and {Tao}, C. and {Masunaga}, K. and {Futaana}, Y. and {Terada}, N.},
        title = "{Inner heliosphere MHD modeling system applicable to space weather forecasting for the other planets}",
      journal = {Space Weather},
         year = 2014,
        month = apr,
       volume = {12},
       number = {4},
        pages = {187-204},
          doi = {10.1002/2013SW000989},
       adsurl = {https://ui.adsabs.harvard.edu/abs/2014SpWea..12..187S}
}

@ARTICLE{Shoda_2018_ApJ_a_self-consistent_model,
       author = {{Shoda}, Munehito and {Yokoyama}, Takaaki and {Suzuki}, Takeru K.},
        title = "{A Self-consistent Model of the Coronal Heating and Solar Wind Acceleration Including Compressible and Incompressible Heating Processes}",
      journal = {\apj},
         year = 2018,
        month = feb,
       volume = {853},
       number = {2},
          eid = {190},
        pages = {190},
          doi = {10.3847/1538-4357/aaa3e1},
archivePrefix = {arXiv},
       eprint = {1712.07760},
 primaryClass = {astro-ph.SR},
       adsurl = {https://ui.adsabs.harvard.edu/abs/2018ApJ...853..190S}
}

@ARTICLE{Shoda_2020_ApJ,
       author = {{Shoda}, Munehito and {Suzuki}, Takeru K. and {Matt}, Sean P. and {Cranmer}, Steven R. and {Vidotto}, Aline A. and {Strugarek}, Antoine and {See}, Victor and {R{\'e}ville}, Victor and {Finley}, Adam J. and {Brun}, Allan Sacha},
        title = "{Alfv{\'e}n-wave-driven Magnetic Rotator Winds from Low-mass Stars. I. Rotation Dependences of Magnetic Braking and Mass-loss Rate}",
      journal = {\apj},
         year = 2020,
        month = jun,
       volume = {896},
       number = {2},
          eid = {123},
        pages = {123},
          doi = {10.3847/1538-4357/ab94bf},
archivePrefix = {arXiv},
       eprint = {2005.09817},
 primaryClass = {astro-ph.SR},
       adsurl = {https://ui.adsabs.harvard.edu/abs/2020ApJ...896..123S}
}

@ARTICLE{Shoda_2021_AA,
       author = {{Shoda}, Munehito and {Takasao}, Shinsuke},
        title = "{Corona and XUV emission modelling of the Sun and Sun-like stars}",
      journal = {\aap},
         year = 2021,
        month = dec,
       volume = {656},
          eid = {A111},
        pages = {A111},
          doi = {10.1051/0004-6361/202141563},
archivePrefix = {arXiv},
       eprint = {2106.08915},
 primaryClass = {astro-ph.SR},
       adsurl = {https://ui.adsabs.harvard.edu/abs/2021A&A...656A.111S}
}

@ARTICLE{Shoda_2022_ApJ,
       author = {{Shoda}, Munehito and {Iwai}, Kazumasa and {Shiota}, Daikou},
        title = "{Testing the Alfv{\'e}n-wave Model of the Solar Wind with Interplanetary Scintillation}",
      journal = {\apj},
         year = 2022,
        month = apr,
       volume = {928},
       number = {2},
          eid = {130},
        pages = {130},
          doi = {10.3847/1538-4357/ac581e},
archivePrefix = {arXiv},
       eprint = {2202.10768},
 primaryClass = {astro-ph.SR},
       adsurl = {https://ui.adsabs.harvard.edu/abs/2022ApJ...928..130S}
}

@ARTICLE{Shoda_2023_ApJ,
       author = {{Shoda}, Munehito and {Cranmer}, Steven R. and {Toriumi}, Shin},
        title = "{Formulating Mass-loss Rates for Sun-like Stars: A Hybrid Model Approach}",
      journal = {\apj},
         year = 2023,
        month = nov,
       volume = {957},
       number = {2},
          eid = {71},
        pages = {71},
          doi = {10.3847/1538-4357/acfa72},
archivePrefix = {arXiv},
       eprint = {2309.09399},
 primaryClass = {astro-ph.SR},
       adsurl = {https://ui.adsabs.harvard.edu/abs/2023ApJ...957...71S}
}

@ARTICLE{Shoda_2024_AA,
       author = {{Shoda}, Munehito and {Namekata}, Kosuke and {Takasao}, Shinsuke},
        title = "{Assessing the capability of a model-based stellar XUV estimation}",
      journal = {\aap},
         year = 2024,
        month = nov,
       volume = {691},
          eid = {A152},
        pages = {A152},
          doi = {10.1051/0004-6361/202450129},
archivePrefix = {arXiv},
       eprint = {2408.16216},
 primaryClass = {astro-ph.SR},
       adsurl = {https://ui.adsabs.harvard.edu/abs/2024A&A...691A.152S}
}

@ARTICLE{Shoda_2025_ApJ,
       author = {{Shoda}, Munehito and {Tokoro}, Kyogo and {Shiota}, Daikou and {Imada}, Shinsuke},
        title = "{Empirical Optimization of the Source-surface Height in the Potential Field Source Surface Extrapolation}",
      journal = {\apj},
         year = 2025,
        month = nov,
       volume = {993},
       number = {2},
          eid = {242},
        pages = {242},
          doi = {10.3847/1538-4357/ae10ba},
archivePrefix = {arXiv},
       eprint = {2510.05513},
 primaryClass = {astro-ph.SR},
       adsurl = {https://ui.adsabs.harvard.edu/abs/2025ApJ...993..242S}
}

@ARTICLE{Shu_1988_JCP,
       author = {{Shu}, Chi-Wang and {Osher}, Stanley},
        title = "{Efficient Implementation of Essentially Non-oscillatory Shock-Capturing Schemes}",
      journal = {Journal of Computational Physics},
         year = 1988,
        month = aug,
       volume = {77},
       number = {2},
        pages = {439-471},
          doi = {10.1016/0021-9991(88)90177-5},
       adsurl = {https://ui.adsabs.harvard.edu/abs/1988JCoPh..77..439S}
}

@ARTICLE{Skirvin_2025_ApJ,
       author = {{Skirvin}, Samuel J. and {Fedun}, Viktor and {Verth}, Gary and {Ballai}, Istvan},
        title = "{3D Magnetohydrodynamic Wave Propagation and Energy Transport in a Simulated Solar Vortex}",
      journal = {\apj},
         year = 2025,
        month = jul,
       volume = {988},
       number = {1},
          eid = {18},
        pages = {18},
          doi = {10.3847/1538-4357/adce73},
archivePrefix = {arXiv},
       eprint = {2504.12745},
 primaryClass = {astro-ph.SR},
       adsurl = {https://ui.adsabs.harvard.edu/abs/2025ApJ...988...18S}
}

@ARTICLE{Sokolov_2013_ApJ,
       author = {{Sokolov}, Igor V. and {van der Holst}, Bart and {Oran}, Rona and {Downs}, Cooper and {Roussev}, Ilia I. and {Jin}, Meng and {Manchester}, IV, Ward B. and {Evans}, Rebekah M. and {Gombosi}, Tamas I.},
        title = "{Magnetohydrodynamic Waves and Coronal Heating: Unifying Empirical and MHD Turbulence Models}",
      journal = {\apj},
         year = 2013,
        month = feb,
       volume = {764},
       number = {1},
          eid = {23},
        pages = {23},
          doi = {10.1088/0004-637X/764/1/23},
archivePrefix = {arXiv},
       eprint = {1208.3141},
 primaryClass = {astro-ph.SR},
       adsurl = {https://ui.adsabs.harvard.edu/abs/2013ApJ...764...23S}
}

@ARTICLE{Solanki_1992_AandA,
       author = {{Solanki}, S.~K. and {Brigljevic}, V.},
        title = "{Continuum brightness of solar magnetic elements}",
      journal = {\aap},
         year = 1992,
        month = sep,
       volume = {262},
       number = {2},
        pages = {L29-L32},
       adsurl = {https://ui.adsabs.harvard.edu/abs/1992A&A...262L..29S}
}

@ARTICLE{Soler_2026_AA,
       author = {{Soler}, Roberto},
        title = "{Alfv{\'e}n wave propagation in the partially ionized lower solar atmosphere: A test of the single-fluid approximation}",
      journal = {\aap},
         year = 2026,
        month = mar,
       volume = {708},
          eid = {A68},
        pages = {A68},
          doi = {10.1051/0004-6361/202659417},
archivePrefix = {arXiv},
       eprint = {2603.04178},
 primaryClass = {astro-ph.SR},
       adsurl = {https://ui.adsabs.harvard.edu/abs/2026A&A...708A..68S}
}

@ARTICLE{Spitzer_1953_PhysRev,
       author = {{Spitzer}, Lyman and {H{\"a}rm}, Richard},
        title = "{Transport Phenomena in a Completely Ionized Gas}",
      journal = {Physical Review},
         year = 1953,
        month = mar,
       volume = {89},
       number = {5},
        pages = {977-981},
          doi = {10.1103/PhysRev.89.977},
       adsurl = {https://ui.adsabs.harvard.edu/abs/1953PhRv...89..977S}
}

@ARTICLE{Stangalini_2014_AandA,
       author = {{Stangalini}, M. and {Consolini}, G. and {Berrilli}, F. and {De Michelis}, P. and {Tozzi}, R.},
        title = "{Observational evidence for buffeting-induced kink waves in solar magnetic elements}",
      journal = {\aap},
         year = 2014,
        month = sep,
       volume = {569},
          eid = {A102},
        pages = {A102},
          doi = {10.1051/0004-6361/201424221},
archivePrefix = {arXiv},
       eprint = {1408.3987},
 primaryClass = {astro-ph.SR},
       adsurl = {https://ui.adsabs.harvard.edu/abs/2014A&A...569A.102S}
}

@ARTICLE{Stansby_2020_JOSS,
       author = {{Stansby}, David and {Yeates}, Anthony and {Badman}, Samuel},
        title = "{pfsspy: A Python package for potential field source surface modelling}",
      journal = {The Journal of Open Source Software},
         year = 2020,
        month = oct,
       volume = {5},
       number = {54},
          eid = {2732},
        pages = {2732},
          doi = {10.21105/joss.02732},
       adsurl = {https://ui.adsabs.harvard.edu/abs/2020JOSS....5.2732S}
}

@ARTICLE{Stansby_2021_AandA,
       author = {{Stansby}, D. and {Ber{\v{c}}i{\v{c}}}, L. and {Matteini}, L. and {Owen}, C.~J. and {French}, R.~J. and {Baker}, D. and {Badman}, S.~T.},
        title = "{Sensitivity of solar wind mass flux to coronal temperature}",
      journal = {\aap},
         year = 2021,
        month = jun,
       volume = {650},
          eid = {L2},
        pages = {L2},
          doi = {10.1051/0004-6361/202039789},
archivePrefix = {arXiv},
       eprint = {2009.13918},
 primaryClass = {astro-ph.SR},
       adsurl = {https://ui.adsabs.harvard.edu/abs/2021A&A...650L...2S}
}

@ARTICLE{Suresh_1997_JCP,
       author = {{Suresh}, A. and {Huynh}, H.~T.},
        title = "{Accurate Monotonicity-Preserving Schemes with Runge Kutta Time Stepping}",
      journal = {Journal of Computational Physics},
         year = 1997,
        month = sep,
       volume = {136},
       number = {1},
        pages = {83-99},
          doi = {10.1006/jcph.1997.5745},
       adsurl = {https://ui.adsabs.harvard.edu/abs/1997JCoPh.136...83S}
}

@ARTICLE{Suzuki_2005_ApJ,
       author = {{Suzuki}, Takeru K. and {Inutsuka}, Shu-ichiro},
        title = "{Making the Corona and the Fast Solar Wind: A Self-consistent Simulation for the Low-Frequency Alfv{\'e}n Waves from the Photosphere to 0.3 AU}",
      journal = {\apjl},
         year = 2005,
        month = oct,
       volume = {632},
       number = {1},
        pages = {L49-L52},
          doi = {10.1086/497536},
archivePrefix = {arXiv},
       eprint = {astro-ph/0506639},
 primaryClass = {astro-ph},
       adsurl = {https://ui.adsabs.harvard.edu/abs/2005ApJ...632L..49S}
}

@ARTICLE{Suzuki_2013_PASJ,
       author = {{Suzuki}, Takeru K. and {Imada}, Shinsuke and {Kataoka}, Ryuho and {Kato}, Yoshiaki and {Matsumoto}, Takuma and {Miyahara}, Hiroko and {Tsuneta}, Saku},
        title = "{Saturation of Stellar Winds from Young Suns}",
      journal = {\pasj},
         year = 2013,
        month = oct,
       volume = {65},
          eid = {98},
        pages = {98},
          doi = {10.1093/pasj/65.5.98},
archivePrefix = {arXiv},
       eprint = {1212.6713},
 primaryClass = {astro-ph.SR},
       adsurl = {https://ui.adsabs.harvard.edu/abs/2013PASJ...65...98S}
}

@ARTICLE{Suzuki_2018_PASJ,
       author = {{Suzuki}, Takeru K.},
        title = "{Stellar winds and coronae of low-mass Population II/III stars}",
      journal = {\pasj},
         year = 2018,
        month = jun,
       volume = {70},
       number = {3},
          eid = {34},
        pages = {34},
          doi = {10.1093/pasj/psy023},
archivePrefix = {arXiv},
       eprint = {1710.04478},
 primaryClass = {astro-ph.SR},
       adsurl = {https://ui.adsabs.harvard.edu/abs/2018PASJ...70...34S}
}

@ARTICLE{Suzuki_2025_PASJ,
       author = {{Suzuki}, Takeru K. and {Ohnaka}, Keiichi and {Yasuda}, Yuki},
        title = "{Red giant winds driven by Alfv{\'e}n waves with magnetic diffusion}",
      journal = {\pasj},
         year = 2025,
        month = aug,
       volume = {77},
       number = {4},
        pages = {863-879},
          doi = {10.1093/pasj/psaf058},
archivePrefix = {arXiv},
       eprint = {2501.00294},
 primaryClass = {astro-ph.SR},
       adsurl = {https://ui.adsabs.harvard.edu/abs/2025PASJ...77..863S}
}

@ARTICLE{Tokoro_2026_ApJ,
       author = {{Tokoro}, Kyogo and {Shoda}, Munehito and {Imada}, Shinsuke},
        title = "{Proposal of a Novel Physical Parameter Characterizing Solar Wind Speed in a Wave-driven Model}",
      journal = {\apj},
         year = 2026,
        month = feb,
       volume = {997},
       number = {2},
          eid = {351},
        pages = {351},
          doi = {10.3847/1538-4357/ae2fea},
archivePrefix = {arXiv},
       eprint = {2601.21229},
 primaryClass = {astro-ph.SR},
       adsurl = {https://ui.adsabs.harvard.edu/abs/2026ApJ...997..351T}
}

@ARTICLE{Tsuneta_2008_ApJ,
       author = {{Tsuneta}, S. and {Ichimoto}, K. and {Katsukawa}, Y. and {Lites}, B.~W. and {Matsuzaki}, K. and {Nagata}, S. and {Orozco Su{\'a}rez}, D. and {Shimizu}, T. and {Shimojo}, M. and {Shine}, R.~A. and {Suematsu}, Y. and {Suzuki}, T.~K. and {Tarbell}, T.~D. and {Title}, A.~M.},
        title = "{The Magnetic Landscape of the Sun's Polar Region}",
      journal = {\apj},
         year = 2008,
        month = dec,
       volume = {688},
       number = {2},
        pages = {1374-1381},
          doi = {10.1086/592226},
archivePrefix = {arXiv},
       eprint = {0807.4631},
 primaryClass = {astro-ph},
       adsurl = {https://ui.adsabs.harvard.edu/abs/2008ApJ...688.1374T}
}

@ARTICLE{Utz_2009_AA,
       author = {{Utz}, D. and {Hanslmeier}, A. and {M{\"o}stl}, C. and {Muller}, R. and {Veronig}, A. and {Muthsam}, H.},
        title = "{The size distribution of magnetic bright points derived from Hinode/SOT observations}",
      journal = {\aap},
         year = 2009,
        month = apr,
       volume = {498},
       number = {1},
        pages = {289-293},
          doi = {10.1051/0004-6361/200810867},
archivePrefix = {arXiv},
       eprint = {0912.2637},
 primaryClass = {astro-ph.SR},
       adsurl = {https://ui.adsabs.harvard.edu/abs/2009A&A...498..289U}
}

@ARTICLE{Utz_2010_AandA,
       author = {{Utz}, D. and {Hanslmeier}, A. and {Muller}, R. and {Veronig}, A. and {Ryb{\'a}k}, J. and {Muthsam}, H.},
        title = "{Dynamics of isolated magnetic bright points derived from Hinode/SOT G-band observations}",
      journal = {\aap},
         year = 2010,
        month = feb,
       volume = {511},
          eid = {A39},
        pages = {A39},
          doi = {10.1051/0004-6361/200913085},
archivePrefix = {arXiv},
       eprint = {0912.1965},
 primaryClass = {astro-ph.SR},
       adsurl = {https://ui.adsabs.harvard.edu/abs/2010A&A...511A..39U}
}

@ARTICLE{Utz_2013_AandA,
       author = {{Utz}, D. and {Jur{\v{c}}{\'a}k}, J. and {Hanslmeier}, A. and {Muller}, R. and {Veronig}, A. and {K{\"u}hner}, O.},
        title = "{Magnetic field strength distribution of magnetic bright points inferred from filtergrams and spectro-polarimetric data}",
      journal = {\aap},
         year = 2013,
        month = jun,
       volume = {554},
          eid = {A65},
        pages = {A65},
          doi = {10.1051/0004-6361/201116894},
archivePrefix = {arXiv},
       eprint = {1304.5508},
 primaryClass = {astro-ph.SR},
       adsurl = {https://ui.adsabs.harvard.edu/abs/2013A&A...554A..65U}
}

@ARTICLE{van_Ballegooijen_2011_ApJ,
       author = {{van Ballegooijen}, A.~A. and {Asgari-Targhi}, M. and {Cranmer}, S.~R. and {DeLuca}, E.~E.},
        title = "{Heating of the Solar Chromosphere and Corona by Alfv{\'e}n Wave Turbulence}",
      journal = {\apj},
         year = 2011,
        month = jul,
       volume = {736},
       number = {1},
          eid = {3},
        pages = {3},
          doi = {10.1088/0004-637X/736/1/3},
archivePrefix = {arXiv},
       eprint = {1105.0402},
 primaryClass = {astro-ph.SR},
       adsurl = {https://ui.adsabs.harvard.edu/abs/2011ApJ...736....3V}
}

@ARTICLE{van_Ballegooijen_2016_ApJ,
       author = {{van Ballegooijen}, A.~A. and {Asgari-Targhi}, M.},
        title = "{Heating and Acceleration of the Fast Solar Wind by Alfv{\'e}n Wave Turbulence}",
      journal = {\apj},
         year = 2016,
        month = apr,
       volume = {821},
       number = {2},
          eid = {106},
        pages = {106},
          doi = {10.3847/0004-637X/821/2/106},
archivePrefix = {arXiv},
       eprint = {1602.06883},
 primaryClass = {astro-ph.SR},
       adsurl = {https://ui.adsabs.harvard.edu/abs/2016ApJ...821..106V}
}

@ARTICLE{van_Ballegooijen_2017_ApJ_coronal_loop,
       author = {{van Ballegooijen}, A.~A. and {Asgari-Targhi}, M. and {Voss}, A.},
        title = "{The Heating of Solar Coronal Loops by Alfv{\'e}n Wave Turbulence}",
      journal = {\apj},
         year = 2017,
        month = nov,
       volume = {849},
       number = {1},
          eid = {46},
        pages = {46},
          doi = {10.3847/1538-4357/aa9118},
archivePrefix = {arXiv},
       eprint = {1710.05074},
 primaryClass = {astro-ph.SR},
       adsurl = {https://ui.adsabs.harvard.edu/abs/2017ApJ...849...46V}
}

@ARTICLE{van_der_Holst_2014_ApJ,
       author = {{van der Holst}, B. and {Sokolov}, I.~V. and {Meng}, X. and {Jin}, M. and {Manchester}, W.~B., IV and {T{\'o}th}, G. and {Gombosi}, T.~I.},
        title = "{Alfv{\'e}n Wave Solar Model (AWSoM): Coronal Heating}",
      journal = {\apj},
         year = 2014,
        month = feb,
       volume = {782},
       number = {2},
          eid = {81},
        pages = {81},
          doi = {10.1088/0004-637X/782/2/81},
archivePrefix = {arXiv},
       eprint = {1311.4093},
 primaryClass = {astro-ph.SR},
       adsurl = {https://ui.adsabs.harvard.edu/abs/2014ApJ...782...81V}
}

@ARTICLE{Van_Doorsselaere_2025_AndA,
       author = {{Van Doorsselaere}, T. and {Sieyra}, M.~V. and {Magyar}, N. and {Goossens}, M. and {Banovi{\'c}}, L.},
        title = "{Uniturbulence and Alfv{\'e}n wave solar model}",
      journal = {\aap},
         year = 2025,
        month = apr,
       volume = {696},
          eid = {A166},
        pages = {A166},
          doi = {10.1051/0004-6361/202450630},
archivePrefix = {arXiv},
       eprint = {2502.14406},
 primaryClass = {astro-ph.SR},
       adsurl = {https://ui.adsabs.harvard.edu/abs/2025A&A...696A.166V}
}

@ARTICLE{Verdini_2007_ApJ,
       author = {{Verdini}, Andrea and {Velli}, Marco},
        title = "{Alfv{\'e}n Waves and Turbulence in the Solar Atmosphere and Solar Wind}",
      journal = {\apj},
         year = 2007,
        month = jun,
       volume = {662},
       number = {1},
        pages = {669-676},
          doi = {10.1086/510710},
archivePrefix = {arXiv},
       eprint = {astro-ph/0702205},
 primaryClass = {astro-ph},
       adsurl = {https://ui.adsabs.harvard.edu/abs/2007ApJ...662..669V}
}

@ARTICLE{Verdini_2019_SolPhys,
       author = {{Verdini}, Andrea and {Grappin}, Roland and {Montagud-Camps}, Victor},
        title = "{Turbulent Heating in the Accelerating Region Using a Multishell Model}",
      journal = {\solphys},
         year = 2019,
        month = may,
       volume = {294},
       number = {5},
          eid = {65},
        pages = {65},
          doi = {10.1007/s11207-019-1458-y},
       adsurl = {https://ui.adsabs.harvard.edu/abs/2019SoPh..294...65V}
}

@ARTICLE{Vidotto_2021_LRSP,
       author = {{Vidotto}, Aline A.},
        title = "{The evolution of the solar wind}",
      journal = {Living Reviews in Solar Physics},
         year = 2021,
        month = dec,
       volume = {18},
       number = {1},
          eid = {3},
        pages = {3},
          doi = {10.1007/s41116-021-00029-w},
archivePrefix = {arXiv},
       eprint = {2103.15748},
 primaryClass = {astro-ph.SR},
       adsurl = {https://ui.adsabs.harvard.edu/abs/2021LRSP...18....3V}
}








\bsp	
\label{lastpage}
\end{document}